\documentclass[superscriptaddress,twocolumn,showpacs,preprintnumbers,prc,floatfix]{revtex4} 
\usepackage{graphicx}
\usepackage{psfrag}
\usepackage{longtable}

\newcommand{\BO}{Helmholtz-Institut f\"ur Strahlen- und Kernphysik, Universit\"at Bonn, D-53115 Bonn, Germany}
\newcommand{\HH}{Institut f\"ur Experimentalphysik, Universit\"at Hamburg, D-22761 Hamburg, Germany}
\newcommand{\JU}{Institut f\"ur Kernphysik, Forschungszentrum J\"ulich, D-52425 J\"ulich, Germany}
\newcommand{\authors}{
  \author{F.~Bauer}        \affiliation{\HH}
  \author{J.~Bisplinghoff} \affiliation{\BO}
  \author{K.~B\"u\ss er}   \affiliation{\HH}
  \author{M.~Busch}        \affiliation{\BO}
  \author{T.~Colberg}      \affiliation{\HH}
  \author{C.~Dahl}         \affiliation{\BO}
  \author{L.~Demir\"ors}   \affiliation{\HH}
  \author{P.D.~Eversheim}  \affiliation{\BO}
  \author{K.O.~Eyser}      \affiliation{\HH}
  \author{O.~Felden}       \affiliation{\JU}
  \author{R.~Gebel}        \affiliation{\JU}
  \author{J.~Greiff}       \affiliation{\HH}
  \author{F.~Hinterberger} \affiliation{\BO}
  \author{E.~Jonas}        \affiliation{\HH}
  \author{H.~Krause}       \affiliation{\HH}
  \author{C.~Lehmann}      \affiliation{\HH}
  \author{J.~Lindlein}     \affiliation{\HH}
  \author{R.~Maier}        \affiliation{\JU}
  \author{A.~Meinerzhagen} \affiliation{\BO}
  \author{C.~Pauly}        \affiliation{\HH}
  \author{D.~Prasuhn}      \affiliation{\JU}
  \author{H.~Rohdje\ss}    \affiliation{\BO}
  \author{D.~Rosendaal}    \affiliation{\BO}
  \author{P.~von Rossen}   \affiliation{\JU}
  \author{N.~Schirm}       \affiliation{\HH}
  \author{W.~Scobel}       \affiliation{\HH}
  \author{K.~Ulbrich}      \affiliation{\BO}
  \author{E.~Weise}        \affiliation{\BO}
  \author{T.~Wolf}         \affiliation{\HH}
  \author{R.~Ziegler}      \affiliation{\BO}
}
\newcommand{\institutes}{}

\begin{document}
\title{Excitation  functions of spin correlation parameters $A_{NN}$,
  $A_{SS}$, and $A_{SL}$ in elastic
  $\stackrel{\rightarrow}{p}\stackrel{\rightarrow}{p}$ scattering
  between 0.45 and  2.5 GeV}
\authors
\institutes
\collaboration{EDDA Collaboration}
\date{Received: date / Revised version: date}
\begin{abstract}
  Excitation functions of the spin correlation coefficients
  $A_{NN}(p_{lab},\theta_{c.m.})$, $A_{SS}(p_{lab},\theta_{c.m.})$,
  and $A_{SL}(p_{lab},\theta_{c.m.})$ have been measured with the
  polarized proton beam of the Cooler Synchrotron COSY and an internal
  polarized atomic beam target.
  Data were taken continuously during the acceleration for proton
  momenta $p_{lab}$ ranging from 1000 to 3300 MeV/c (kinetic energies
  $T_{lab}$ 450 - 2500 MeV) as well as for discrete momenta of 1430
  MeV/c and above 1950 MeV/c covering angles $\theta_{c.m.}$ between
  30$^{\circ}$ and 90$^{\circ}$.
  The data are of high internal consistency.
  Whereas $A_{SL}(p_{lab,}\theta_{c.m.})$ is small and without
  structures in the whole range, $A_{NN}$ and even more $A_{SS}$ show
  a pronounced energy dependence.
  The angular distributions for $A_{SS}$ are at variance with
  predictions of existing phase shift analyses at energies beyond 800
  MeV.
  The impact of our results on phase shift solutions is discussed.
  The direct reconstruction of the scattering amplitudes from all
  available {\sl pp} elastic scattering data considerably reduces the
  ambiguities of solutions.
\end{abstract}

\maketitle

\section{Introduction}
\label{sec:10}
This paper reports on the final part of a major experimental program
devoted to a precision measurement of proton-proton elastic scattering
by using the polarized beam of the Cooler Synchrotron COSY in
conjunction with a polarized atomic beam target.

The EDDA experiment \cite{ALB97,ALT00,EDD03} has
been conceived to provide highly accurate data of internal 
consistency for many projectile energies between 0.45 and 2.5 GeV
covering an angular range in $\theta_{c.m.}$ from 30$^{\circ}$ to
90$^{\circ}$.
For this purpose, it has been set up as internal beam experiment.
Elastically scattered  protons are detected in coincidence by a
cylindrical multi-layered scintillator hodoscope. Data acquisition
occurs during beam acceleration to measure quasi-continuous excitation
functions as it was first done at SATURNE \cite{GAR85}.
A highly polarized atomic hydrogen beam is used as target for fast and
easy spin manipulation with magnetic guide fields to minimize
systematic errors, a technique extensively applied by the PINTEX
collaboration at IUCF \cite{VPR98,RAT98} at energies below 500 MeV.

Nucleon-nucleon (NN) interaction is a process fundamental to the
understanding of the nuclear forces between free nucleons as well as
in the nuclear environment.
Elastic NN scattering data, condensed into energy dependent solutions
of phase-shift analyses (PSA) \cite{BYS90,STO93,NAG96,ARN97,ARN00},
are used as an important ingredient in theoretical calculations
modelling  nuclear interactions.
Below the pion production threshold at about 280 MeV elastic NN
scattering is described with impressive precision \cite{MAC01} by
several approaches, e.g. modern phenomenological and meson theoretical
models \cite{LAC80,MAC87,STO94,WIR95,MAC01a}, and more recently chiral
perturbation theory \cite{BED02}.

Up to 800~MeV sufficient data exist that still allow an unambiguous
determination of phase shift parameters and that are reasonably well
reproduced by extended meson exchange models \cite{EYS04}.
For even higher energies the number of contributing partial waves
increases, and at the same time are the data more scarce and
inconsistent. As an example no data are available for $A_{SS}$ between 
$T_{lab} =$ 792~MeV and 5~GeV.
This coefficient is particularly sensitive to the spin-spin and
spin-tensor parts of the NN interaction and the corresponding
scattering amplitudes \cite{EDD03}.
This may be one reason for the serious discrepancies between the PSA
solutions of different groups \cite{ARN00,BYS98} in the regime
$T_{lab} >$ 1.2~GeV, that could not be resolved with the (model
independent) direct reconstruction of the scattering amplitudes.
The final part of the EDDA experiment therefore aims at a substantial
improvement of the data base on observables for the scattering of
polarized protons on polarized protons.

In the first phase of the EDDA experiment, thin poly\-pro\-py\-lene
$(CH_2)_n$ fibers have been used in the circulating COSY beam to
determine excitation functions of unpolarized differential cross
sections \cite{ALB97,EDD04}.
These data prompted a considerable modification and extension of PSA
solutions up to 2.5 GeV \cite{ARN97}.
In the second phase it was continued \cite{ALT00,EDD05} with the
unpolarized COSY beam impinging on the polarized atomic beam target to 
access excitation functions of the analyzing power
$A_N(p_{lab},\theta_{c.m.})$.
In addition the results for $A_N$ are an important ingredient  for a
consistent analysis of the double polarized experiment presented here,
because they allow to fix the overall polarization scale.

A short account of the results for the correlation coefficients of the 
third phase has been given in \cite{EDD03}, 
where their angular distributions were presented for the projectile
energy 2.11 GeV.
It was observed that the existing PSA  solutions \cite{ARN00,BYS98}
are in sharp contrast to the observable $A_{SS}$.
The direct reconstruction of the scattering amplitudes (DRSA) with
inclusion of our results helped to reduce ambiguities in the
scattering amplitudes, indicating that these coefficients indeed
provide additional constraints to the extraction of scattering
amplitudes and phase shifts.

Here  we present excitation functions $A_{NN}(p_{lab},\theta_{c.m.})$, 
$A_{SS}(p_{lab},\theta_{c.m.})$, and $A_{SL}(p_{lab},\theta_{c.m.})$
from measurements during the projectile beam acceleration as well as
for 10 fixed energies ranging from 0.772 GeV to 2.493 GeV.
They are compared to existing PSA solutions and enter into additional
DRSA wherever the accumulated data base allows.
Many details of the experiment and its analysis have been discussed in
\cite{EDD04,EDD05}, to which we refer the reader for additional
information.
Here we concentrate on  aspects of the experiment and its analysis for
the double polarized
$\stackrel{\rightarrow}{p}\stackrel{\rightarrow}{p}$ case.
The paper is accordingly organized as follows: In Sec.\ \ref{sec:20} we
give a short account of the experimental setup and the measurements
performed.
Sec.\ \ref{sec:24} deals with the background reduction and selection of
valid scattering events.
The data analysis is described in Sec.\ \ref{sec:30} with emphasis on
the determination of asymmetries, polarizations, correlation
coefficients, and the minimization of their systematic errors.
The results are then presented as excitation functions and angular
distributions in Sec.\ \ref{sec:5}, followed by a DRSA for five
projectile energies.

\section{The experiment}
\label{sec:20}
\subsection{Detector and target setup}
\label{sec:21}

\begin{figure}
  \centerline{\includegraphics[width=8.8cm]{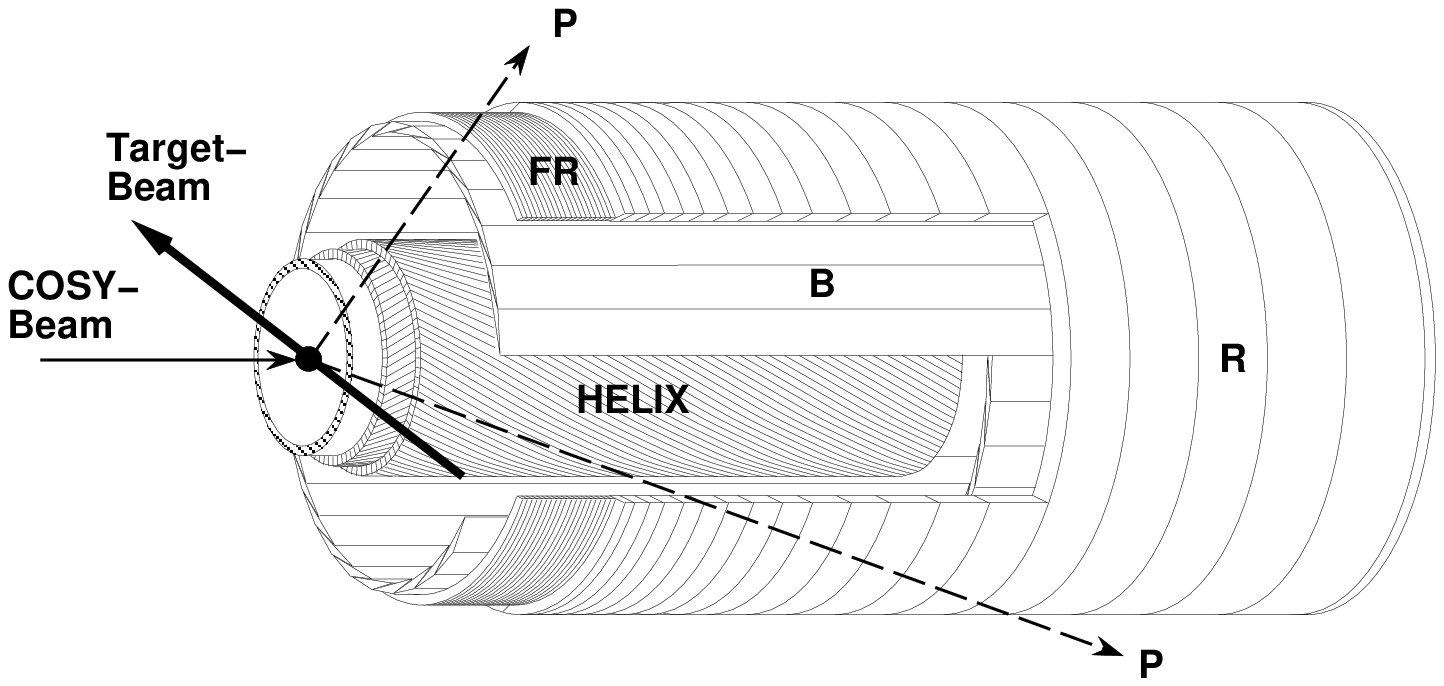}}
  \centerline{\includegraphics[width=8.8cm]{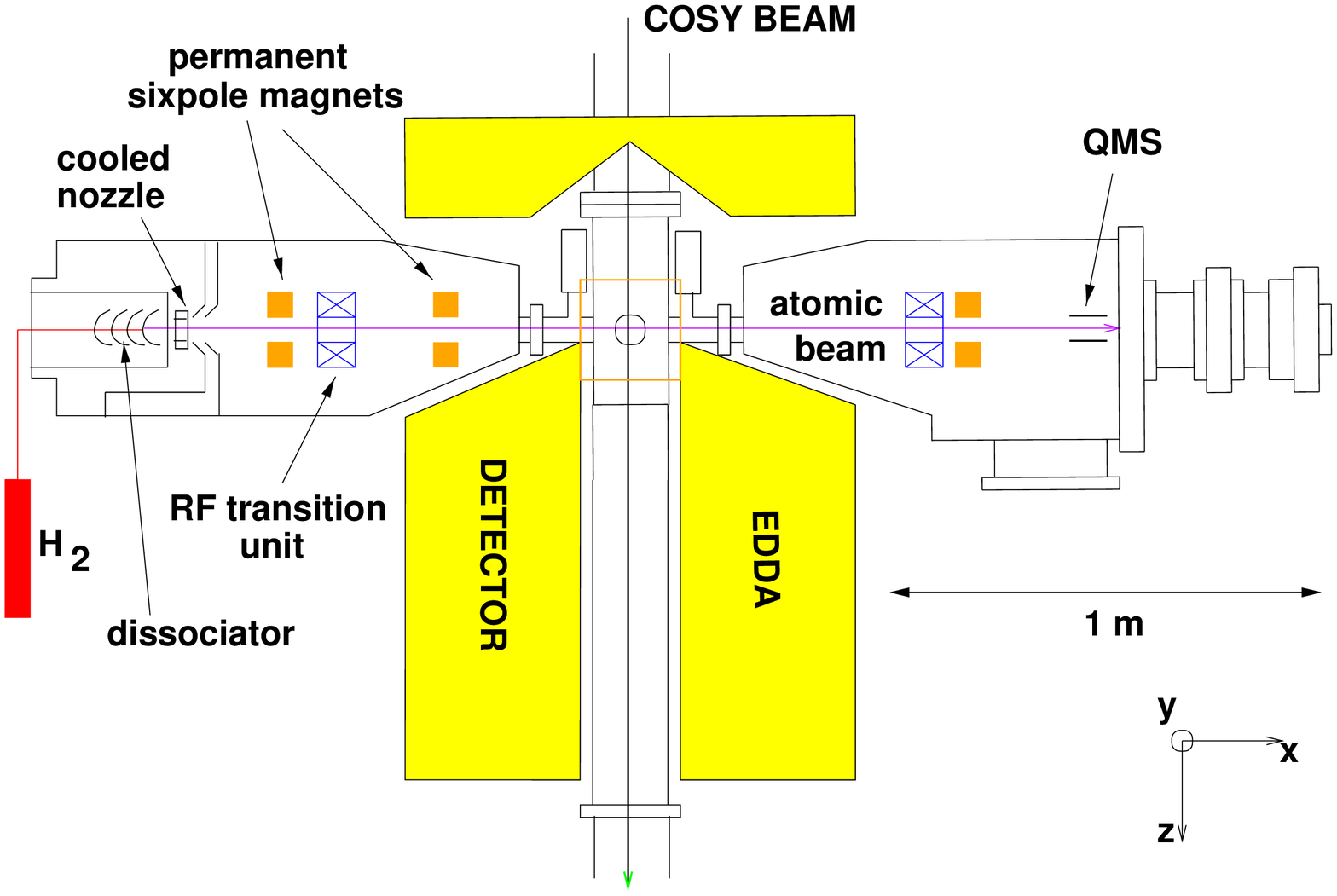}}
  \caption{Scheme of the EDDA detector (top) and its combination with the atomic beam target (bottom).}
  \label{fig:1}
\end{figure}

The detector shown schematically in Fig.\ \ref{fig:1} consists 
of two cylindrical shells covering 30$^{\circ}$ to 150$^{\circ}$
in $\theta_{c.m.}$ for the elastic {\sl pp} channel and
about 85\% of the full solid angle.
The inner shell (HELIX) is composed of 4 layers of 160 scintillating fibers  
which are helically wound in opposing directions.
The outer shell consists of 32 scintillator bars (B) which are running parallel
to the beam axis. They are surrounded by 29 scintillator rings (R; FR), split 
into left and right semirings to allow independent radial readout of the 
scintillation light. 
The scintillator cross sections were designed in such a way  that
each particle traversing the outer layers produces a
signal in two neighbouring bars and rings. Analysis of the
fractional light output is used to improve the polar and azimuthal FWHM 
angle resolution to about 1$^{\circ}$ and
1.9$^{\circ}$, respectively. This geometry allows for a vertex
reconstruction with a resolution of about 1 mm in the $x$-, $y$- and
$z$-direction.

The polarized target \cite{EVE97}  is shown in Fig.\ \ref{fig:1}, too. Hydrogen 
atoms with nuclear polarization are prepared in an atomic beam source
with dissociator, cooled nozzle, permanent sixpole magnets, and RF-transition units, 
where the former remove one of the two electron spin states and the latter induce a 
transition to a thus unpopulated hyperfine state, with only one nuclear spin state remaining. 
This preparation provides an atomic beam of $\sim$12 mm width (FWHM) and up to 2$\cdot 10^{11}$ 
$H$ atoms/cm$^2$ areal density at the 
intersection with the COSY beam, and a peak polarization of 90\%. Details of the 
target performance and polarization distribution are given in \cite{EDD05}. 

\begin{figure}
  \psfrag{Ferritjoch}{ferrite yoke}
  \psfrag{Targetstutzen}{flange}
  \psfrag{Strahlrohr}{beam pipe}
  \centerline{\includegraphics[width=5cm]{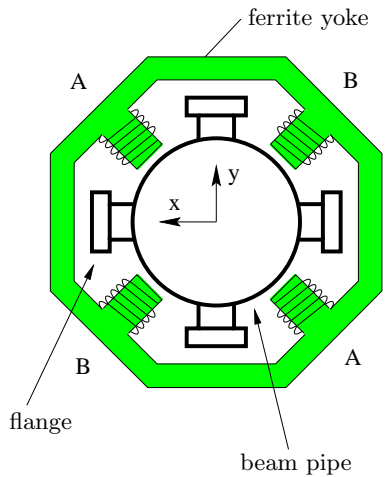}}
  \caption{Magnet configuration generating the guide field in the
    $xy$-plane.
    The yoke carries 4 pole shoes that produce the field components in
    diagonal direction and surrounds the beam pipe.
    The atomic beam crosses the beam pipe horizontally.}
  \label{fig:2}
\end{figure}

The direction of the target polarization in the vertex volume is defined by a magnetic guide field. Its 
components in the $xy$-plane are generated with two pairs of dipole magnets (A and B in Fig.\ \ref{fig:2}) 
arranged at $z = 0$ in the $xy$-plane under $\pm$45$^{\circ}$ and  $\pm$135$^{\circ}$.  
Superposition of their fields of same strength yields components $\pm B_x$ or $\pm B_y$ depending on the 
polarities applied to the two pairs. The magnets are equipped with ferrite yokes such that field strengths 
in the order of 1 mT can be achieved with moderate, easily switchable
currents (5 A). They exceed ambient 
field components by almost two orders of magnitude and thus guide the spin direction reliably. On the 
other hand distortions of the orbiting protons are sufficiently small; the angular kicks result in 
momentum dependent horizontal and vertical shifts between 20 and 50 $\mu$m.
Components $\pm B_z$ are achieved with two solenoids mounted concentric to the COSY beam line 
upstream and downstream of the nominal target position.    

\subsection{COSY beam}
\label{sec:22}
$H^-$ ions are preaccelerated to $T_{lab}$ = 45 MeV with high nuclear
polarization ($\ge$ 80\%) normal to the storage orbit plane
($y$-direction) and are then stripping injected into the COSY storage
ring.
The protons are further accelerated with a ramping speed of 1.15
(GeV/c)/s to one of the ten flattop values $T_{ft}$ of 0.772, 1.226,
1.358, 1.546, 1.800, 1.939, 2.110, 2.301, 2.377, and 2.493 GeV with
typically $3\cdot10^9$ - $1.5\cdot10^{10}$ protons circulating.

The momentary energies were derived from the RF of the cavities and
the circumference of the closed orbit with uncertainties increasing 
from 0.25 to 2 MeV with energy. The reconstruction of beam parameters is described in \cite{EDD04}; 
they vary with the momentary energy, but remain constant from cycle to cycle. COSY was tuned in a way 
that in vertical ($y$) direction the beam centroid and profile (6 mm FWHM) were not dependent on the 
momentary energy; as a consequence, the effective target polarization resulting for the overlap 
region with the 12 mm wide atomic beam remains constant during the ramping.  

\begin{figure}
  \psfrag{Polarisation}{polarization}
  \psfrag{Strahlimpuls (MeV/c)}{momentum (MeV/c)}
  \centerline{\includegraphics[width=8.8cm]{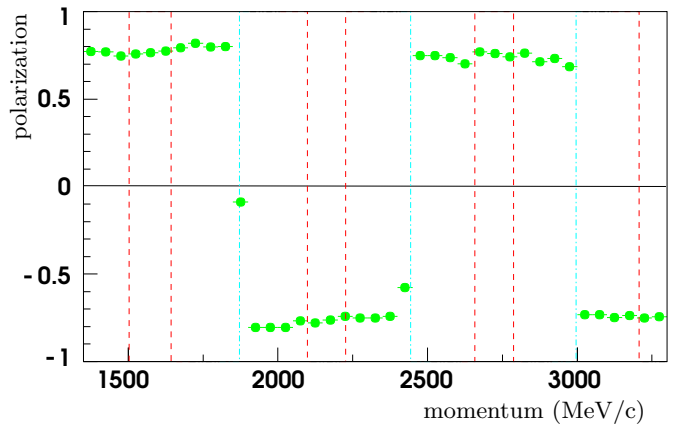}}
  \caption{Depolarization of beam protons during acceleration across imperfection and 
    intrinsic resonances indicated by vertical lines (the former with
    spin flip) to the flattop momentum of 3.3 GeV/c (kinetic energy 2.5 GeV).}
  \label{fig:3}
\end{figure}

During acceleration the spins of the stored, polarized  protons  precess around the direction of 
the COSY guide fields normal to the orbit and experience depolarizing resonances. The so called 
imperfection resonances occur, if there is an integer number of precessions per turn such that 
field components in the orbit plane give rise to coherently accumulating distortions. In addition depolarization 
can  be due to intrinsic resonances excited by horizontal field components from vertical focusing 
that cause betatron oscillations around the  nominal orbit. At COSY, techniques have been developed 
\cite{LEH03}  to cross both types of resonances, partly under spin flip, with a minimum of polarization 
loss. Figure \ref{fig:3} demonstrates the preservation of polarization during acceleration to the 
highest flattop energy as it was measured with the EDDA detector being operated as internal 
polarimeter \cite{LEH03}. 

\subsection{Measurements}
\label{sec:23}
The excitation functions $A_{NN}(p_{lab},\theta_{c.m.})$,
$A_{SS}(p_{lab},\theta_{c.m.})$, and $A_{SL}(p_{lab},\theta_{c.m.})$
were simultaneously measured in a sequence of acceleration
cycles.
Data acquisition started during ramping at 1 GeV/c (0.45 GeV) and
extended over the flattop of 6 s length  before the beam was
decelerated to complete a COSY cycle by returning to the injection
status after 13 s.
Typical luminosities per cycle were 1.0 - 4.0$\cdot 10^{27}$ cm$^{-2}$s$^{-1}$.
Sufficient statistics for excitation functions covering the full
energy range from 0.45 GeV onward was achieved by accumulation of data 
in over 6$\cdot 10^5$  such cycles with an integrated luminosity of 12
nb$^{-1}$.
The direction of the target polarization in the vertex volume was changed from cycle to cycle by switching  
the magnetic guide field in a sequence $+x, -x, +y, -y, +z, -z$ that was then repeated with the beam 
polarization flipped from $+y$ to $-y$. Such supercycles including 12 accelerator cycles were formed in 
order to minimize systematic errors in the extraction of the correlation coefficients (cf.\ Sec.\ 
\ref{sec:32}) due to long term drifts of beam and/or target properties.    

Measurements were performed in four running periods of up to 7 weeks length each. Each period was devoted to 2 - 4 
flattop energies, with slightly varying conditions as to luminosity, cycle timing, maximum polarizations, 
and background  conditions. Altogether 4.6$\cdot 10^6$ events were taken during ramping and 
12.5$\cdot 10^6$ in the flattop time periods.

\section{Data Reconstruction}
\label{sec:24}
\subsection{Selection of elastic events}
\label{sec:241}
The on-line triggering and off-line identification of elastic {\sl pp} scattering is based on the requirement 
for coplanarity
\begin{equation}
\varphi_1 - \varphi_2 = 180^{\circ}
\label{eqn0.1}
\end{equation}
and for kinematic correlation
\begin{equation}
\tan\theta_1\cdot\tan\theta_2=2\cdot\frac{m_p\cdot c^2}{(2\cdot m_p\cdot c^2+T_{lab})}
\label{eqn0.2}
\end{equation}
with $\theta_i$ and $\varphi_i$ denoting polar and azimuthal angle of the proton $i$ in the laboratory system, 
$m_p$ their mass and $T_{lab}$ the projectile proton energy. The geometry and granularity of the outer 
scintillator shell enables for two-prong events a fast trigger on these two requirements. 

In the off-line analysis the trajectories of these correlated prongs
are reconstructed from the hit and timing pattern in the inner and
outer detector shell.
The vertex associated with the trajectories is determined
geometrically as the point of their closest approach in the target
region.
It is obtained with a FWHM resolution of 1.3 mm in $x$ and $y$ and 0.9
mm in $z$.
The scattering angles $\theta_i, \varphi_i$ are calculated with
respect to this vertex position and transformed in the center-of-mass
(c.m.) system assuming the kinematics of elastic {\sl pp} scattering.
The resulting angular resolution is 1.4$^{\circ}$ in $\theta_{c.m.}$
and 1.9$^{\circ}$ in $\varphi$.

Momentum conservation then requires the trajectories of elastic {\sl pp} scattering to fulfill a 
180$^{\circ}$ correlation in the c.m. frame. The spatial angle deviation from this 
back-to-back scattering, furtheron 
referred to as {\it kinematic deficit} $\alpha$, can originate from finite angular resolution and angular 
straggling. It will, however, also occur for the vast majority of nonelastic background events, that  
can therefore be substantially suppressed with a cut on $\alpha$. The cut was optimized on data 
with known composition of elastic and inelastic events from our event
generator \cite{ACK02}, and unpolarized EDDA-data \cite{EDD04} (using $(CH_2)_n$ and 
carbon fiber targets), leading to a momentum dependence, viz. 
\begin{equation}
\alpha \le \alpha_{max}(p_{lab}) =(8.32 - 0.72\cdot\frac{p_{lab}}{1~{\rm GeV/c}})^{\circ}.
\label{eqn0.3}
\end{equation}

The basic geometrical trajectory and vertex reconstruction is supplemented by a vertex fit. It improves 
the reconstruction within the limits of the spatial and angular resolution under the constraints 
of elastic scattering kinematics with intersecting trajectories. In case of convergence the $\chi^2_{vert}$ 
of this fit can be used as additional criterion for event selection.

\subsection{Background reduction}
\label{sec:242}
Inelastic reactions and scattering involving unpolarized protons are
sources of background and should be reduced or well known in the
analysis. 
The detector is a pure hodoscope and does not allow for particle
identification.
Elastic events produce two sets of piercing points in both the inner and the
outer detector shells.
The hit pattern in the outer shell comprises two scintillator bars (B) and one
semi-ring (R; FR) in each of the left and right sides.
In the inner shell (HELIX) four scintillating fibers can be combined to two
piercing points.
Crosstalking between neighbouring channels increases the number of
accepted fibers to six.
The hit pattern selection reduces the amount of data by a factor of
2.
Further analysis is then based on a converging vertex fit.
The momentum dependent cut on the kinematic deficit, eq.\ (\ref{eqn0.3}),
removes another 5\% from the reconstructed events and restrains inelastic
events to less than 1\% in the remaining data.

Reconstructed vertices can occur far off the overlap region of
projectile beam and atomic beam target, especially in the direction of
the COSY beam.
These events are outside the magnetic guide field region and comprise
reactions with residual gas.
This leads to a decreased beam polarization, which is suppressed by a
cut on the $z$-vertex: -15~mm $\leq z\leq$ 20~mm.
Similar effects arise in the $xy$-plane and are avoided by an
elliptical cut with the axes being taken as 3 times the widths
$\sigma_x$ and $\sigma_y$ of momentum dependent vertex distributions
(cf.\ \cite{EDD04,EDD05}).

After all cuts applied no more than 6\% of the collected data remain
for the determination of spin correlation coefficients.

\section{Data analysis}
\label{sec:30}
\subsection{Nomenclature and coordinates}
\label{sec:31}

\begin{figure}
  \psfrag{Beschleunigerebene}{detector plane}
  \psfrag{Streuebene}{scattering plane}
  \centerline{\includegraphics[width=8.8cm]{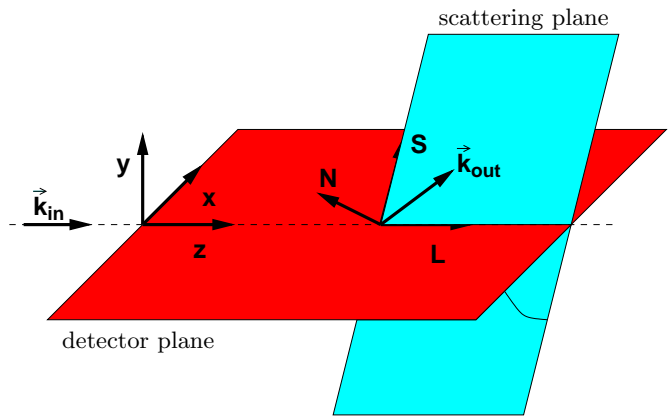}}
  \caption{Coordinate systems: detector (laboratory) and scattering frames.}
  \label{fig:4}
\end{figure}

Polarization observables are described here by attaching a frame of
reference to the projectile (and target) proton following the Madison
convention (\cite{BYS78}, cf.\ Fig.\ \ref{fig:4}).
Its momenta $\vec{k}_{in}$ and 
$\vec{k}_{out}$ define the scattering plane, and  $N$ is normal to it; 
 $L$ points in the direction of $\vec{k}_{in}$, and  $S$ 
completes the right handed frame. Using the Argonne notation \cite{BOU80} the differential 
cross section for scattering projectile protons of polarization $\vec{P}$ on target 
protons of polarization $\vec{Q}$ is then given by
\begin{eqnarray}
\frac{d\sigma(\theta,\varphi)}{d\Omega}/I_0=&1&
+A_N\;\;\,(P_N+Q_N)\nonumber\\ 
&&+A_{NN}\;P_NQ_N \nonumber\\ 
&&+A_{SS}\;\,P_SQ_S\nonumber\\
&&+A_{SL}(P_SQ_L+P_LQ_S)\nonumber\\ 
&&+A_{LL}\;\,P_LQ_L.
\label{eq1}
\end{eqnarray}
Here, $I_0=\left(\frac{d\sigma(\theta)}{d\Omega}\right)_0$ denotes
the unpolarized differential cross section.
In the experiment, $\vec{P}$ and $\vec{Q}$ are expressed in the frame
$x$, $y$,  $z$ that refers to the horizontal plane of the nominal
projectile trajectory and the symmetry axis of the EDDA detector.
It is transformed into the scattering frame with a rotation
around the beam axis by the azimuthal angle $\varphi$.
At present COSY provides only protons with polarization
$\vec{P}=(0,P_y,0)$ such that eq.\ (\ref{eq1}) yields 
\begin{eqnarray}
\frac{d\sigma(\theta,\varphi)}{d\Omega}/I_0&=1&
+A_N\;\;\,[(P_y+Q_y)\cos\varphi-Q_x\sin\varphi]\nonumber \\
&&+A_{NN}[P_yQ_y\cos^2\varphi-P_yQ_x\sin\varphi\cos\varphi]\nonumber\\
&&+A_{SS}\;\,[P_yQ_y\sin^2\varphi+P_yQ_x\sin\varphi\cos\varphi]\nonumber\\
&&+A_{SL}\;\;P_yQ_z\sin\varphi.
\label{eq2}
\end{eqnarray}

The polarization observables $A_N$, $A_{SS}$, $A_{NN}$, and $A_{SL}$
can be deduced from the azimuthal modulation of the polarized cross
section if the polarizations $P_y, Q_x, Q_y$, and $Q_z$ are known. 

For an unpolarized beam, $\vec{P}$ = 0, and a target polarization
$Q_y$  eq.\ (\ref{eq2}) reduces to
\begin{equation}
\frac{d\sigma(\theta,\varphi)}{d\Omega}=I_0\cdot 
(1+A_N\cdot Q_y\cdot\cos\varphi).
\label{eq3}
\end{equation}

\subsection{Determination of spin correlation coefficients}
\label{sec:32}
The number  $N(\theta, \varphi, \vec{P}, \vec{Q})$ of scattering events is related to the coefficients via 
\begin{equation}
N(\theta, \varphi, \vec{P}, \vec{Q}) = \frac{d\sigma(\theta, \varphi)}{d\Omega}\cdot \Delta\Omega \cdot 
L(\vec{P}, \vec{Q})\cdot \eta(\theta, \varphi)
\label{eq5}
\end{equation}
with the integrated luminosity $L$, the detection efficiency $\eta$, and the solid angle $\Delta\Omega$ 
subtended by the detector element.

In \cite{ALT00,EDD05} the analyzing power $A_N$ has been obtained by calculating the azimuthal asymmetry from 
the numbers of events 
for scattering to the left $[N_L(\theta)]$ and the right side $[N_R(\theta)]$. In order to correct for false 
asymmetries \cite{OHL73}, measurements were performed with opposite polarizations $Q_{+y}$ and $Q_{-y}$ to 
determine the geometrical means $R(\theta) = \sqrt{N_{L_-}(\theta)N_{R_+}(\theta)}$ and 
$L(\theta) = \sqrt{N_{L_+}(\theta)N_{R_-}(\theta)}$. Starting from
eq.\ (\ref{eq3}), the left-right asymmetry 
$\epsilon_{LR} = (L(\theta) - R(\theta))/(L(\theta) + R(\theta))$ allows to calculate $A_N$ from
\begin{equation}
A_N \langle\cos{\varphi}\rangle = \frac{\epsilon_{LR}}{Q_y}
\label{eq4}
\end{equation}
for identical detector segments centered around the azimuthal positions $\varphi$ and $\varphi + \pi$ and 
$\langle\cos{\varphi}\rangle$ being the weight\-ed mean for a segment. Similarly, $A_N$ was calculated from the 
runs with horizontal polarization $Q_{\pm x}$,  the bottom-top asymmetry 
$\epsilon_{BT}(\theta, \varphi)$ and the 
weighted mean $<\sin{\varphi}>$. Details are given in \cite{ALT00,EDD05}. 

The coefficients $A_{NN}, A_{SS}$, and $A_{SL}$ can be extracted in a similar way, however  with 
asymmetries that constitute an extension of the formalism applied to deduce $A_N$. For this purpose, 
the azimuthal coverage of the detector is subdivided into 4 identical segments centered around 
$\varphi = \frac{\pi}{4}, \frac{3\pi}{4}, \frac{5\pi}{4}$, and $\frac{7\pi}{4}$. The respective numbers
of events are denoted by $N^n$, with $n = 1, 3, 5$, or $7$. They vary with the orientation of the 
polarizations $P_y$ and $Q_i$ ($i$ = x, y, z), which are therefore indicated as subscripts, e.g. as 
$N^{3}_{+-}$ in case of polarizations $+P_y$ and $-Q_i$. For each quadrant and value $i$ there are 4 
numbers of events  ($N^i_{++}, N^i_{-+}, N^i_{--}, N^i_{+-}$), which yield 48 numbers of events for 
the 12 different polarization combinations.   
  
Inspection of eq.\ (\ref{eq2}) reveals, that each 4 out of the 16 numbers of events for a given target 
polarization $Q_i$ represent the same cross section (e.g. for $Q_x: N^1_{++}, N^3_{-+}, N^5_{--}, N^7_{+-}$) 
and can be combined to geometrical mean values 
$N_1(Q_x) = (N^1_{++}\cdot N^3_{-+}\cdot N^5_{--}\cdot N^7_{+-})^{\frac{1}{4}}$, 
$N_2(Q_x) = (N^1_{+-}\cdot N^3_{--}\cdot N^5_{-+}\cdot N^7_{++})^{\frac{1}{4}}$, 
$N_3(Q_x) = (N^1_{-+}\cdot N^3_{++}\cdot N^5_{+-}\cdot N^7_{--})^{\frac{1}{4}}$, and 
$N_4(Q_x) = (N^1_{--}\cdot N^3_{+-}\cdot N^5_{++}\cdot N^7_{-+})^{\frac{1}{4}}$.
Similar combinations are found \cite{BAU01} for $Q_y$ and $Q_z$ . This way the 16 numbers of events are 
reduced to 4 such mean values, which are then used to define 3 different
asymmetries for each of the three target polarizations $Q_i$:
\begin{eqnarray}
\epsilon_1(Q_i) &=& \frac{N_1(Q_i)+N_2(Q_i)-N_3(Q_i)-N_4(Q_i)}{N_1(Q_i)+N_2(Q_i)+N_3(Q_i)+N_4(Q_i)},\nonumber \\
\epsilon_2(Q_i) &=& \frac{N_1(Q_i)-N_2(Q_i)+N_3(Q_i)-N_4(Q_i)}{N_1(Q_i)+N_2(Q_i)+N_3(Q_i)+N_4(Q_i)},\\
\epsilon_3(Q_i) &=& \frac{N_1(Q_i)-N_2(Q_i)-N_3(Q_i)+N_4(Q_i)}{N_1(Q_i)+N_2(Q_i)+N_3(Q_i)+N_4(Q_i)}.\nonumber
\label{eq6}
\end{eqnarray}

Evaluation of the 9 asymmetries $\epsilon_1(Q_x)$,
$\dots\epsilon_3(Q_z)$ with eq.\ (\ref{eq2}) leads to the 
following expressions 

\begin{eqnarray}
  \epsilon_1(Q_x)&=&P_y\cdot A_N\langle\cos{\varphi}\rangle,
  \label{eq7}\\
  \epsilon_2(Q_x)&=&-Q_x\cdot A_N\langle\sin{\varphi}\rangle,
  \label{eq8}\\
  \epsilon_3(Q_x)&=&P_y\cdot Q_x\cdot(A_{SS}-A_{NN})\langle\sin{\varphi}\cos{\varphi}\rangle,
  \label{eq9}\\
  \epsilon_1(Q_y)&=&P_y\cdot A_N\langle\cos{\varphi}\rangle,
  \label{eq10}\\
  \epsilon_2(Q_y)&=&Q_y\cdot A_N\langle\cos{\varphi}\rangle,
  \label{eq11}\\
  \epsilon_3(Q_y)&=&P_y\cdot Q_y\cdot(A_{SS}\langle\sin^2{\varphi}\rangle+A_{NN}\langle\cos^2{\varphi}\rangle),
  \label{eq12}\\
  \epsilon_1(Q_z)&=&P_y\cdot A_N\langle\cos{\varphi}\rangle,
  \label{eq13}\\
  \epsilon_2(Q_z)&=&0,
  \label{eq14}\\
  \epsilon_3(Q_z)&=&P_y\cdot Q_z\cdot A_{SL}\langle\sin{\varphi}\rangle.
  \label{eq15}
\end{eqnarray}
With the analyzing power $A_N$ being known from \cite{ALT00,EDD05},  the average value $P$ of the beam 
polarization $P_y$ is derived from eqs.\ (\ref{eq7}), (\ref{eq10}), and (\ref{eq13}). Target 
polarizations $Q_x$ and $Q_y$ are obtained from eqs.\ (\ref{eq8}) and (\ref{eq11}); the average value 
$Q$ is used for $Q_z$ as well, because the polarized atomic beam is aligned with the magnetic guide  
field in the interaction zone, a process not correlated with the generation of polarization in the 
atomic beam source. The remaining eqs.\ (\ref{eq9}), (\ref{eq12}), and
(\ref{eq15}) allow then to determine $A_{NN}$, $A_{SS}$, and $A_{SL}$
from the respective asymmetries including the polarizations $P$ and $Q$.

\subsubsection{Corrections of asymmetries}
\label{sec:321}
Eqs.\ (\ref{eq7}) - (\ref{eq15}) are based on the assumption, that the
detector efficiencies do not change between measurements with flipped
polarizations and are the same for the four azimuthal segments.
Changing efficiencies would lead to false asymmetries and wrong
geometrical mean values.
The numbers of events can be efficiency-corrected, though.
The sum of all events from the possible polarization combinations
comprise an unpolarized measurement with no azimuthal dependence
except for efficiency differences.
To correct for the efficiency the calculated expectation values of the
trigonometric functions are replaced by means that apply the real
numbers of events for weighting. These weighting factors were Gaussian  
distributed with typical standard deviations of 8\% - 10\%; they constitute 
also an additional correction of other false asymmetries.

Knowledge of the COSY beam intensity is not necessary, as long as there are no
systematical differences between parallel and antiparallel beam and target
polarizations, namely $\pm Q_y$.
Integral beam intensities have been measured for all polarization combinations
and have been used for correction of the numbers of events.

\subsection{Systematic errors}
\label{sec:33}
\begin{figure}
  \psfrag{ass}{$A_{SS}$}
  \psfrag{ann}{$A_{NN}$}
  \psfrag{asl}{$A_{SL}$}
  \psfrag{theta}{\hspace{-0.8cm}$\theta_{c.m.}~(deg)$}
  \centerline{\includegraphics[width=8.8cm]{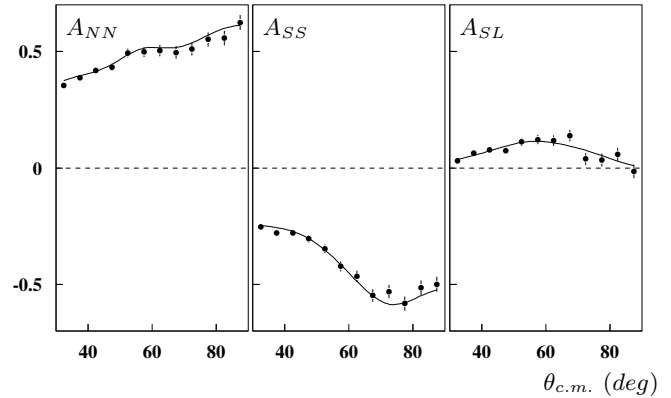}}
  \caption{Correlation coefficients $A_{SS}$ (left side), $A_{NN}$ (middle), $A_{SL}$ (right) as obtained 
    from an analysis of $12\cdot 10^6$ elastic events that were simulated for $T_{lab}$ = 1546 MeV, $P$ = 0.8 and 
    $Q$ = 0.7 with an angular distribution given by the SAID solution FA00 (solid lines).}
  \label{fig:5}
\end{figure}

In a first step we have checked the analysis scheme outlined by applying it to Monte Carlo generated
events. The simulation was developed for and applied to the measurement of 
the unpolarized \cite{ALB97} excitation functions and those of the analyzing power \cite{ALT00,EDD05}. It 
includes the detector geometry in all details, energy deposition of charged reaction products, their 
hadronic and electromagnetic interaction in the detector material. The event generator is described 
in \cite{ACK02}; it produces the elastic part of the input in accordance with 
the solution FA00 of the phase shift analysis of \cite{ARN00}. Data analysis occurs with the same 
tools that are applied to real data. Typical polarization values $P = 0.8$ and 
$Q = 0.7$ were used to generate elastic events at $T_{lab}$ = 1546 MeV. Their analysis reproduced these 
polarizations ($P =0.804\pm0.004$ and $Q = 0.703\pm0.006$) as well as the spin correlation 
coefficients (cf.\ Fig.\ \ref{fig:5}) essentially within the statistical uncertainties and thus 
confirmed the scheme culminating in eqs.\ (\ref{eq7}) - (\ref{eq15}).

There are, however, several sources of possible systematic errors that are associated with 
deviations of the real polarization scenario from the simulated one, or with possible correlations 
of the polarizations to other quantities entering into eq.\ (\ref{eq4}). Those which may have a sizeable 
impact on the analysis will be discussed in some detail.

\subsubsection{Misalignment of polarizations $\vec{Q}$ and  $\vec{P}$}
\label{sec:331}
\begin{figure}
  \psfrag{Ass}{$A_{SS}$}
  \psfrag{Ann}{$A_{NN}$}
  \psfrag{Asl}{$A_{SL}$}
  \psfrag{THETACM}{\hspace{-.7cm}$\theta_{c.m.}~(deg)$}
  \centerline{\includegraphics[width=8.8cm]{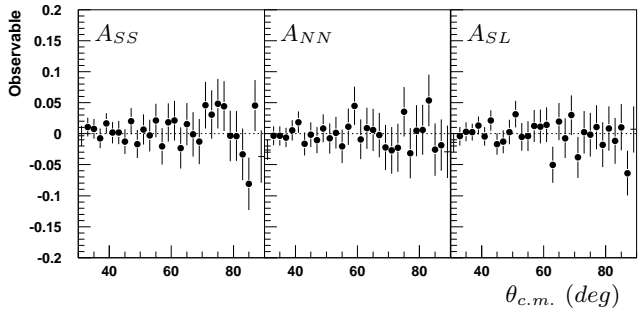}}
  \caption{Deviations $\Delta A_{SS}$ (left side), $\Delta A_{NN}$ (middle), $\Delta A_{SL}$ (right) of 
    the correlation coefficients from the values of Fig.\ \ref{fig:4} obtained with additional constant 
    polarization components $\delta Q_x$ = 0.05 and $\delta Q_y$ = 0.05 in addition to the main component 
    $Q$.}
  \label{fig:6}
\end{figure}

Target polarizations $\vec{Q}$ may deviate in the interaction region from the intended direction due to (i) 
a misalignment of the guide field $\vec{B}$ or (ii) additional external field components not sufficiently 
compensated. In case (i) additional polarization components are generated that change their directions 
together with a reversion of the guide field. In contrast, (ii) causes a constant field component not 
sensitive to a flip of the guide field. These two cases have therefore been studied separately 
\cite{BAU01}. Insertion of a main component $Q_x$ (or $Q_y$) with additional small 
components $\delta Q_z$ and $\delta Q_y$ (or $\delta Q_x$) into eq.\ (\ref{eq2}) yields false 
asymmetries that depend for (ii) quadratically on $\delta Q$, because all first order terms cancel through 
the formation of geometrical mean values $N(Q)$. False asymmetries are therefore expected to be small. 
Monte Carlo simulations indeed show no systematic deviations within
the statistical uncertainties, as can be seen in the example in Fig.\
\ref{fig:6} for the case of additional, constant components.
The same 
results hold for components that flip with the main component, although the dependence on 
$\delta Q_z$ is in this case linear. The resulting false asymmetry, however, is proportional to $A_{SL}$ 
(see eq.\ (\ref{eq15})), and this coefficient is generally small compared to $A_{SS}$ and $A_{NN}$. 

It remains to be shown that the deviating components of the guide
field $\vec{B}$ in the interaction region are indeed sufficiently
small.
For this purpose, simultaneous measurements of $B_x, B_y$, and $B_z$
have been performed with a fluxgate sensor (Bartington MAG-03MCTP).
It allowed to scan $\vec{B}$ in steps of 5 mm in three dimensions with
a dynamical range from $10^{-9}$ T to $10^{-3}$ T.
In the vertex region permanent residual $\vec{B}$ components were
observed with absolute values in the order of $10^{-5}$ T; they were
compensated by offset values of the guide field coils.
The main components of the guide field were typically 0.7$\cdot
10^{-3}$ T.
Field gradients perpendicular to its 
nominal direction gave rise to additional components of up to 2$\cdot
10^{-5}$ T; they generate maximum deviations from the nominal
directions of a main guide field $B_z$ $(B_x, B_y)$ of less than
3.5$^{\circ}$ (1.5$^{\circ}$) in the fiducial interaction volume.
This is small compared to the deviations assumed for the Monte Carlo
calculations.
The resulting errors of $Q$ components are therefore estimated to be
less than 0.2\%.

Deviations $\delta P$ of the absolute beam polarization may occur
with revision of the polarization direction from $+P$ to $-P$ as $|\pm
P| = P \pm \delta P$; they are, however, eliminated by the geometrical
mean values $N(P)$ of the numbers of events in first order, such that
only $(\delta P)^2$ terms enter into eqs.\ (\ref{eq6}).
As a consequence, simulated deviations $\delta P$ up to absolute 
values $\pm$0.05 have negligible impact on correlation coefficients
$A_{SS}$, $A_{NN}$, $A_{SL}$ or polarizations $P, Q$.
The same result is obtained for deviations $\delta Q$.
Moreover, the generation of the beam polarization and the alignment of
the spins along the $x$-, $y$-, and $z$-directions are independent
processes and therefore $\delta Q$ is expected to vanish.
This has been confirmed in a dedicated analysis of representative
experimental data with standard $\chi^2$ minimization techniques
applied to the set of eqs.\ (\ref{eq2}) for the 12 spin combinations.

\subsubsection{Further systematic errors}
\label{sec:332}
Sources for further systematic errors include unpolarized and inelastic
background.
The unpolarized background was reduced through restrictions of the accepted
vertex region, as described in Sec.\ \ref{sec:242}.
This of course also leads to a loss of polarized scattering events but
improves the effective polarizations and results in decreased
statistical uncertainties of the spin correlation coefficients.
Their values are not affected.

\begin{figure}
  \psfrag{D1}{$\Delta A_{NN}$}
  \psfrag{D2}{$\Delta A_{SS}$}
  \psfrag{momentum}{\hspace{-.0cm}momentum bin}
  \psfrag{theta}{\hspace{-.0cm}$\theta_{c.m.}~(deg)$}
  \centerline{\includegraphics[width=8.8cm]{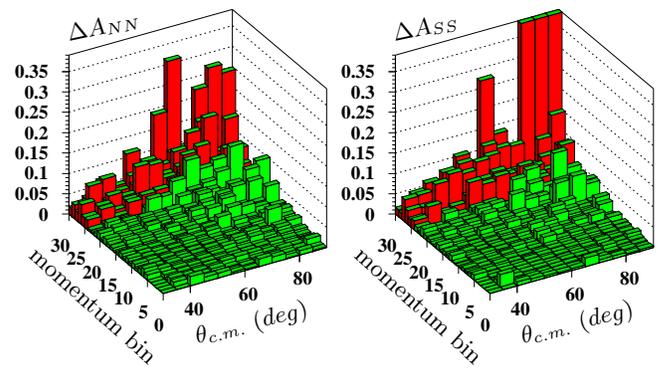}}
  \caption{Maximum deviations $\Delta A_{NN}(p_{lab},\theta_{c.m.})$ 
    and $\Delta A_{SS}(p_{lab},\theta_{c.m.})$ of correlation coefficients after variation of accepted 
    inelastic background for the momentum bins $\Delta p_{lab} =$ 61~MeV/c ranging from 1060 to 2890~MeV/c. 
    Bins beyond 2500~MeV/c are shown in dark.} 
  \label{fig:Deltasinramp}
\end{figure}

\begin{figure}
 \psfrag{diff1}{\hspace{-.8cm}$\Delta A_{NN}$}
  \psfrag{diff2}{\hspace{-.8cm}$\Delta A_{SS}$}
  \psfrag{theta}{\hspace{-1.cm}$\theta_{c.m.}~(deg)$}
  \psfrag{39}{\hspace{.2cm}$2300~MeV/c$}
  \psfrag{89}{\hspace{.2cm}$3180~MeV/c$}
  \psfrag{syst}{$\Delta_{syst}$}
  \psfrag{stat}{$\sigma_{stat}$}
  \centerline{\includegraphics[width=8.8cm]{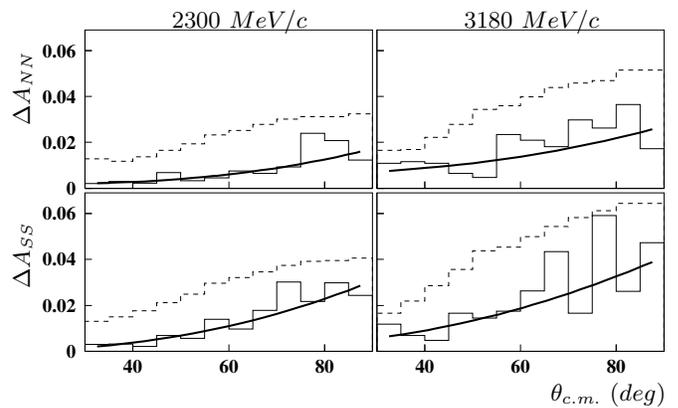}}
  \caption{Comparison of maximum deviations $\Delta A_{NN}$ and $\Delta A_{SS}$ 
    (solid histograms) of correlation coefficients after variation of accepted inelastic background for the 
    flattop measurements at $T_{ft}$ = 1.546~GeV ($p_{lab}$ = 2300~MeV/c) and 2.377~GeV  
    ($p_{lab}$ = 3180~MeV/c) with statistical uncertainties $\sigma_{A_{NN}}$ and $\sigma_{A_{SS}}$ 
    (dashed lines). The thick solid lines are polynomial fits to these maximum deviations.} 
  \label{fig:background}
\end{figure}

The inelastic background is more problematic to access, because there 
are few data of differential cross sections from inelastic reactions
available for Monte Carlo applications.
This leads only to a rough knowledge of the fraction of numbers of
inelastic events and says nothing about their spin dependent
behaviour. 
On the other hand, the effect of the inelastic background can be
estimated directly from the measurement without knowledge of its exact 
fraction.
For this purpose the fraction of accepted inelastic events has been
varied by modifying $\alpha_{max}$ in eq.\ (\ref{eqn0.3}) in small
steps within reasonable limits, and the variations of the spin
correlation coefficients have been deduced.
These variations are highly sensitive to the covered statistics.
Figure \ref{fig:Deltasinramp} shows the maximum deviations for $A_{NN}$
and $A_{SS}$ derived from the data taken during ramping.
On the average they increase with $p_{lab}$ and $\theta_{c.m.}$.
The same procedure was performed with the flattop data.
The results in Fig.\ \ref{fig:background} for two of the flattop
energies demonstrate, that the inelastic background leads to
variations $\Delta A_{ij}$ of less than 0.01 - 0.06 in all three correlation 
coefficients, which is usually less than the statistical
uncertainties. Our error estimates are based on the polynomial fit values. 
We conclude from the comparison of  Fig.\ \ref{fig:Deltasinramp} with
Fig.\ \ref{fig:background} that significant results can be obtained
from the excitation functions below 2500~MeV/c.
For higher momenta flattop data will be preferred and the excitation
function data of this region is excluded from the final results.

\subsection{Consistency checks}
\label{sec:34}
\begin{figure}
  \psfrag{AN}{\hspace{.2cm}$A_N$}
  \psfrag{1765}{1765~MeV}
  \psfrag{1793}{1793~MeV}
  \psfrag{theta}{\hspace{-1.cm}$\theta_{c.m.}~(deg)$}
  \centerline{\includegraphics[width=8.8cm]{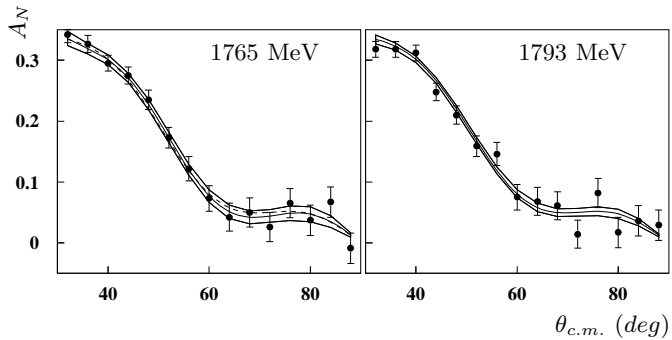}}
  \caption{Angular distributions $A_N(\theta_{c.m.})$ (solid dots) for two projectile energies together 
    with the polynomial best fits and the respective errors; the best fit to the right distribution is 
    repeated as dashed line in the left one as an indication for the momentum dependence.}
  \label{fig:7}
\end{figure}

The analyzing powers $A_N(p_{lab}, \theta_{c.m.})$ entering into eqs.\ 
(\ref{eq7}) - (\ref{eq15}) 
are taken from the preceding stage of the EDDA experiment performed with our polarized atomic beam 
target and the unpolarized COSY beam \cite{ALT00,EDD05}. For this application they have been fitted 
with Legendre polynomial expansions up to $5^{th}$ order and momentum dependent coefficients.
Figure \ref{fig:7} shows an example at medium momenta.

\begin{figure}
  \psfrag{an}{$A_N$}
  \psfrag{3}{$\theta_{c.m.}=40^o$}
  \psfrag{8}{$\theta_{c.m.}=60^o$}
  \psfrag{13}{$\theta_{c.m.}=80^o$}
  \psfrag{p}{\hspace{-2.7cm}momentum (MeV/c)}
  \centerline{\includegraphics[width=8.8cm]{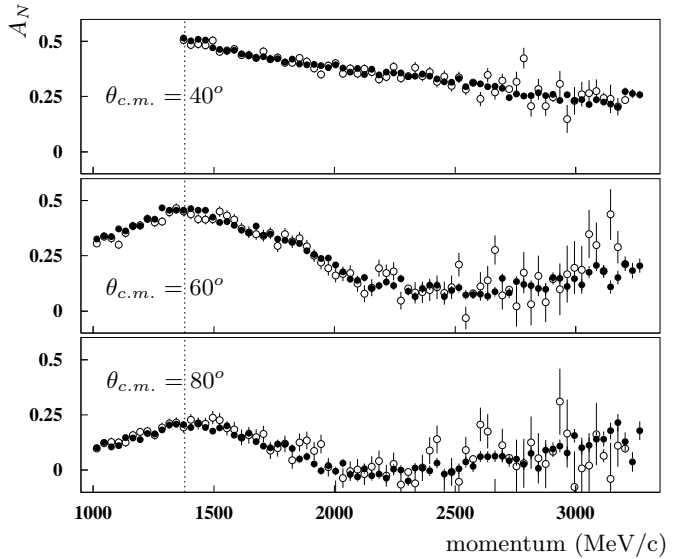}}
  \caption{Excitation functions $A_N(p_{lab}, \theta_{c.m.})$ as obtained from the present experiment (open 
    symbols) and from the single polarization experiment \cite{EDD05} (solid dots). The dashed line indicates 
    the normalization point of the latter to the reference \cite{MCN90}.}
  \label{fig:8}
\end{figure}

In principle the analyzing powers can be derived directly from the
present data, too, by discarding measurements with $Q_{\pm z}$ and
averaging the beam polarization $P_{\pm y}$.
This has been done and some representative excitation functions are
compared in Fig.\ \ref{fig:8} to those of \cite{EDD05}.
The values $A_N$ deduced this way scatter around the statistically
much more precise results of the dedicated $A_N$ experiment, but they
do not indicate systematic deviations.
A quantitative comparison of all excitation functions for
$\theta_{c.m.}$ ranging in increments of 4$^{\circ}$ from 32$^{\circ}$ 
to 88$^{\circ}$ yields reduced values $\chi^{2}_{red}$ between 0.71
and 1.53 with a $\chi^{2}_{red}=$ 0.93 for the whole data set.
This internal consistency is important, because the precise derivation
of the polarizations $P$, $Q$ is based upon it.

The four running periods (cf.\ Sec.\ \ref{sec:23}) contribute with comparable statistics to the 
excitation functions; they differ, however, in several technical aspects. Therefore they were first 
analyzed independently and separately for each of their flattop energies $T_{ft}$. Before merging two such 
subsets $j$ and $k$ to one ensemble of data, their mutual consistency has been checked with a 
$\chi^2$- test
\begin{equation}
\chi^2_{red}=\frac{1}{N-1}\sum_{i=1}^N\frac{(O_i^{(j)}-O_i^{(k)})^2}{\sigma_j^2+\sigma_k^2},
\label{eq16}
\end{equation}
where $O_i^{(j)}(p_{lab},\theta_{c.m.})$ is a spin observable deduced from the $j^{th}$ subset, 
$\sigma_j$ its statistical error, with $i$ running over all $N$ observables common to both subsets. 
The resulting $\chi^2_{red}$ values vary between 0.96 and 2.52 and give no need to discard any of the subsets. 
Therefore all data were combined into one set. 

In a similar way the compatibility of observables from data collected in the flattop times with 
those from the corresponding momentum bin of the combined excitation functions can be checked. 
We find $\chi^2_{red}<$ 1.75 in all cases. The flattop results, due to their small statistical 
uncertainties, therefore complement the excitation functions at high energies in a very consistent 
manner.

\subsection{Error summary}
\label{sec:36}
Estimates for the systematic errors of $A_{NN}$, $A_{SS}$, and
$A_{SL}$ include the contributions from the misalignment of
polarization ($\leq$ 0.01), from incomplete spin flipping ($\leq$
0.01), and from the inelastic background ($\leq$ 0.06); they are
typically smaller than the statistical uncertainties even for flattop
energies, and even more so during ramping.

Normalization uncertainties of the polarizations $P\cdot Q$
arise from the statistical uncertainties of the used $A_{N}$ and
of the measurement of the asymmetries (cf.\ eqs.\
\ref{eq7}-\ref{eq15}).
The resulting uncertainty is raised by the beam polarization
during acceleration, as the target polarization remains constant.
The beam polarization is treated as constant only between depolarizing
resonances and leads to momentum dependent normalization uncertainties 
between 1.1\% and 2.5\% below 2500 MeV/c.
Flattop measurements yield comparable normalization uncertainties
ranging from 2.1\% at 1430 MeV/c up to 4.5\% at 3100 MeV/c (and 2.8\%
at 3300 MeV/c) due to a restricted statistical accuracy of the
determined polarizations.
Additionally the analyzing powers $A_N$ carry an overall absolute
normalization uncertainty of 1.2\% \cite{EDD05}, that spread into the polarizations $P$ and $Q$ 
(cf. eqs. 10, 11, 13, 14, 16) and give rise to a momentum independent 
normalization uncertainty of 1.7\% via eqs. 12, 15, 18 of all spin correlation coefficients.

In the figures representing data of this work, only the statistical
uncertainties are given as error bars.
The systematic errors are listed in the data tables \cite{WWW04}. Here, only the results 
obtained at the flattop energies are tabulated (Table I).

\section{Results and discussion}
\label{sec:5}
\subsection{Excitation functions}
\label{sec:51}
\begin{figure}[t]
  \psfrag{37.5}{$\theta_{c.m.}=37.5^o$} \psfrag{47.5}{$\theta_{c.m.}=47.5^o$}
  \psfrag{57.5}{$\theta_{c.m.}=57.5^o$} \psfrag{67.5}{$\theta_{c.m.}=67.5^o$}
  \psfrag{77.5}{$\theta_{c.m.}=77.5^o$} \psfrag{87.5}{$\theta_{c.m.}=87.5^o$}
  \psfrag{T}{\hspace{-.7cm}$T_{lab}$ (MeV)}
  \psfrag{p}{\hspace{-.9cm}$p_{lab}$ (MeV/c)}
  \psfrag{ramp}{\small excitation function}
  \psfrag{flattop}{\small fixed momentum}
  \psfrag{said}{\small \sc Said Fa00}
  \psfrag{lampf}{\small \sc Lampf}
  \psfrag{saturne}{\small Saturne}
  \psfrag{pnpi}{\small \sc Pnpi}
  \psfrag{sin}{\small \sc Sin}
  \psfrag{anl}{\small \sc Anl}
  \centerline{\includegraphics[width=8.8cm]{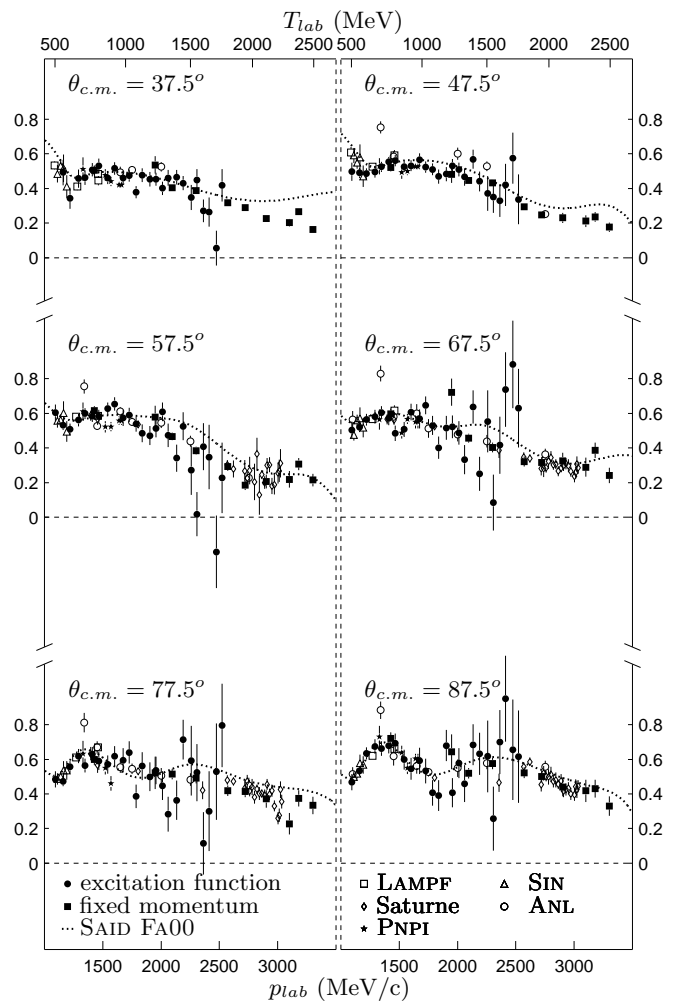}}
  \caption{Excitation functions $A_{NN}(p_{lab})$ for 6 
    angles $\theta_{c.m.}$ together with data from the SAID
    database.
    }
  \label{fig:9}
\end{figure}

\begin{figure}[t]
  \psfrag{37.5}{$\theta_{c.m.}=37.5^o$} \psfrag{47.5}{$\theta_{c.m.}=47.5^o$}
  \psfrag{57.5}{$\theta_{c.m.}=57.5^o$} \psfrag{67.5}{$\theta_{c.m.}=67.5^o$}
  \psfrag{77.5}{$\theta_{c.m.}=77.5^o$} \psfrag{87.5}{$\theta_{c.m.}=87.5^o$}
  \psfrag{T}{\hspace{-.7cm}$T_{lab}$ (MeV)}
  \psfrag{p}{\hspace{-.9cm}$p_{lab}$ (MeV/c)}
  \psfrag{ramp}{excitation function}
  \psfrag{flattop}{fixed momentum}
  \psfrag{said}{\sc Said Fa00}
  \psfrag{lampf}{\sc Lampf}
  \psfrag{sin}{\sc Sin}
  \centerline{\includegraphics[width=8.8cm]{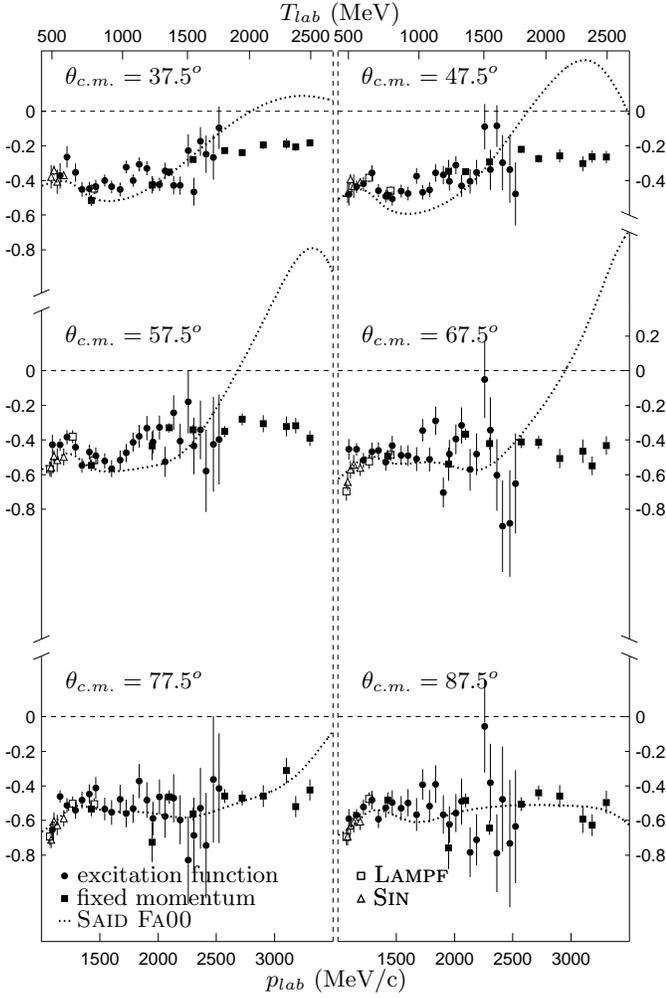}}
  \caption{Same as in Fig.\ \ref{fig:9} for $A_{SS}(p_{lab})$.}
  \label{fig:10}
\end{figure}

\begin{figure}[t]
  \psfrag{37.5}{$\theta_{c.m.}=37.5^o$} \psfrag{47.5}{$\theta_{c.m.}=47.5^o$}
  \psfrag{57.5}{$\theta_{c.m.}=57.5^o$} \psfrag{67.5}{$\theta_{c.m.}=67.5^o$}
  \psfrag{77.5}{$\theta_{c.m.}=77.5^o$} \psfrag{87.5}{$\theta_{c.m.}=87.5^o$}
  \psfrag{T}{\hspace{-.7cm}$T_{lab}$ (MeV)}
  \psfrag{p}{\hspace{-.9cm}$p_{lab}$ (MeV/c)}
  \psfrag{ramp}{excitation function}
  \psfrag{flattop}{fixed momentum}
  \psfrag{said}{\sc Said Fa00}
  \psfrag{lampf}{\sc Lampf}
  \psfrag{saturne}{Saturne}
  \psfrag{sin}{\sc Sin}
  \psfrag{anl}{\sc Anl}
  \centerline{\includegraphics[width=8.8cm]{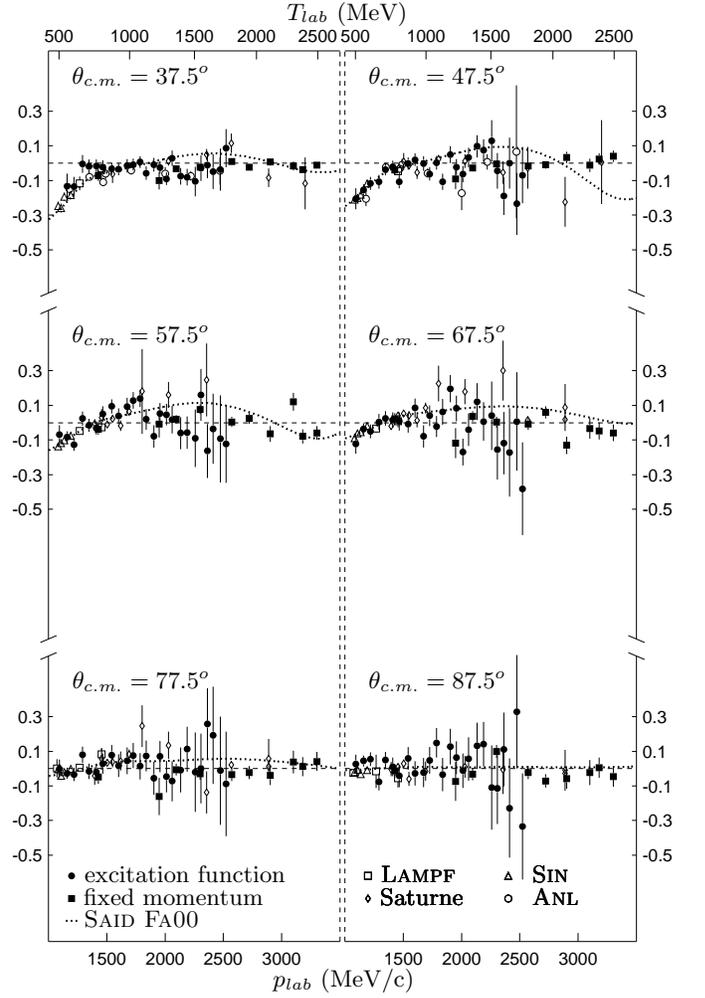}}
  \caption{Same as in Fig.\ \ref{fig:9} for $A_{SL}(p_{lab})$.}
  \label{fig:11}
\end{figure}

The results will be first presented as excitation functions. For this
purpose the data taken during ramping are binned into $\Delta
p_{lab}\approx$ 60~MeV/c (dependent on the position of the
depolarizing resonances) and $\Delta\theta_{c.m.}=$ 5$^{\circ}$
intervals, the latter centered around twelve angles $\theta_{c.m.}$
from 32.5$^{\circ}$ to 87.5$^{\circ}$.
They are supplemented by the data taken at the ten flattop energies
$T_{ft}$.
A representative subset of 18 (out of 36) excitation functions is
shown in Figs.\ \ref{fig:9} - \ref{fig:11}.
For comparison data from other experiments and a global phase shift
solution from fall 2000 \cite{ARN00} have been included in the
figures.

$A_{NN}$ is positive in the whole angular and momentum range and
slowly decreasing with momentum.
Our data fill gaps in the existing data base especially at
intermediate energies and are otherwise in good consistency with other 
measurements \cite{BEL80,MCN81,BHA81,BYS85,LEH87,LES88,BAL99,ALL00,ALL01}.
There are deviations from the PSA solution above
$p_{lab}=2000$ MeV/c, though.
Significant structures at large polar angles at small momenta are
reproduced in the data as well as in the PSA solution.
  
$A_{SS}$ is a crucial observable, as it has so far only been measured below 
$T_{lab} =$ 792~MeV, and for the two energies $T_p =$ 5.1~GeV \cite{AUE76} 
and 10.8~GeV \cite{AUE86} that are beyond the range of present PSA solutions. 
Our data are negative in the covered angle and momentum range (as are the 
high energy data just mentioned) and  in good agreement with measurement of 
\cite{APR83,DIT84} below 792 MeV.
The PSA solution is determined through other observables and becomes
radically different with increasing momentum at medium angles.
While the data are almost momentum independent, the PSA solution rises
after a small drop and even becomes positive above $p_{lab}=$2500
MeV/c. A change of sign cannot be seen in the data at all.

The correlation coefficient $A_{SL}$ is compatible with zero over a wide range of 
energies and angles; only at small angles for momenta below 1400 MeV/c $A_{SL}$, 
i.e. the single spin flip mechanism, has some systematic influence on the 
scattering process. Our data are in general agreement with 
existing \cite{LES88,AUE83,PER88,FON89,ALL98a,GLA92} data, they are,
however, for small momenta and for the fixed momenta mostly superior
in statistics.

\subsection{Angular distributions}
\label{sec:52}
\begin{figure}[t]
  \psfrag{theta}{\hspace{-1cm}$\theta_{c.m.}(deg)$}
  \psfrag{ann}{\hspace{-.2cm}$A_{NN}$}
  \psfrag{ass}{\hspace{-.2cm}$A_{SS}$}
  \psfrag{asl}{\hspace{-.2cm}$A_{SL}$}
  \psfrag{this}{{\sc Edda}, this work}
  \psfrag{saturne}{\sc Saturne}
  \centerline{\includegraphics[width=8.8cm]{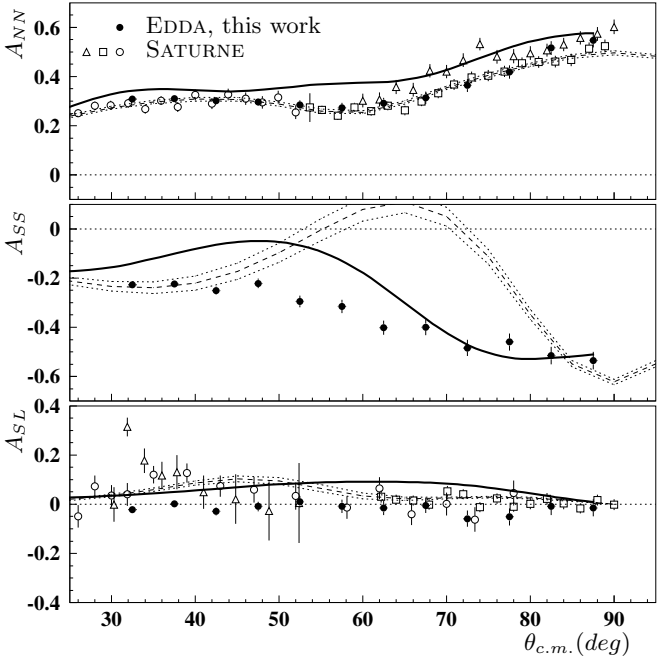}}
  \caption{Angular distributions of spin correlation coefficients $A_{NN}$, $A_{SS}$, and $A_{SL}$ at 
    $T_{lab}$ = 1.8~GeV ($p_{lab}$ = 2572~MeV/c) from this work (solid symbols) and 
    Refs.\ \cite{PER88,FON89,ALL98a,ALL00,ALL01} 
    (open symbols) in comparison to phase shift predictions of SAID (SM00, solid line, Ref.\ \cite{ARN00}) 
    and Bystricky et al. (dashed line with error corridor, Ref.\ \cite{BYS98}).}
  \label{fig:12}
\end{figure}

Rearrangement of the data yields angular distributions for each of the 
24 momentum bins and 10 flattop energies.
In Fig.\ \ref{fig:12} we present the results for $p_{ft}=$ 2572~MeV/c
($T_{ft}=$ 1.8~GeV).
For $A_{NN}$, good agreement is found with the SATURNE data
\cite{LEH87,ALL00,ALL01} and with the PSA solution from \cite{BYS98}
for this fixed energy.
The energy dependent global solution SM00 reproduces the angular
dependence well, but with absolute values being about 20\% above the
experimental ones.
The angular distributions $A_{SL}(\theta_{c.m.})$ turn out to be flat,
as predicted by both phase shift solutions.
However, we cannot confirm the positive values found in \cite{FON89} for
small angles. 

For $A_{SS}$, no data exist to compare with. The PSA solutions therefore essentially represent  
extrapolations beyond the energy 792~MeV; both are in striking disagreement (the agreement at 
90$^{\circ}$ is forced by the identity \cite{BYS78} $A_{SS}=A_{NN}-1-A_{LL}$ with experimental 
data being available for the right hand side). Huge discrepancies like those between the PSA 
solutions in Fig.\ \ref{fig:12} have been observed for the 2.1~GeV data \cite{EDD03}, too, although 
there the single energy solution from \cite{BYS98} shows the larger deviation from our experimental 
data. In \cite{ARN00} these discrepancies were attributed to differences in some partial wave 
solutions. They may reflect a non-uniqueness also visible in the DRSA of \cite{BYS98}. It is therefore 
expected that the addition of the spin correlation coefficients from this work will help to 
remove some of the ambiguities inherent to PSA and DRSA solutions.  

\subsection{Direct reconstruction of scattering amplitudes}
\label{sec:53}
Knowledge of the scattering amplitudes uniquely defines the phase
shifts and all observables of nucleon nucleon scattering.
The transition matrix $T$ for elastic $\stackrel{\rightarrow}{p}\stackrel{\rightarrow}{p}$ 
scattering is fully determined by five complex amplitudes \cite{BYS78}.
Using the positive and negative helicity states $\mid+\rangle$ and
$\mid-\rangle$ in the c.m. frame, these helicity amplitudes are \cite{JAC59}:
\begin{eqnarray}
  \phi_1=\langle++\mid{\bf T}\mid++\rangle, &\quad&
  \phi_4=\langle+-\mid{\bf T}\mid-+\rangle, \nonumber\\ 
  \phi_2=\langle++\mid{\bf T}\mid--\rangle, &\quad&
  \phi_5=\langle++\mid{\bf T}\mid+-\rangle, \label{eqn:helicity}\\
  \phi_3=\langle+-\mid{\bf T}\mid+-\rangle. &\quad& \nonumber
\end{eqnarray}
Obviously the helicity amplitudes are directly connected to the NN-interaction with its dependence 
on double ($\phi_2, \phi_4$), single ($\phi_5$), and no ($\phi_1, \phi_3$) spin flip.  All observables, 
and in particular the correlation coefficients of this paper, can be expressed by these amplitudes:
\begin{eqnarray}
  A_{SS}\cdot I_0&=&Re(\phi_1\phi_2^\star+\phi_3\phi_4^\star),
\label{eq21}\\
  A_{NN}\cdot I_0&=&Re(\phi_1\phi_2^\star-\phi_3\phi_4^\star)+2|\phi_5|^2,
\label{eq22}\\
  A_{SL}\cdot I_0&=&Re([\phi_1+\phi_2-\phi_3+\phi_4]\phi_5^\star)
\label{eq23}
\end{eqnarray}
and can thus be related to the different kinds of spin dependence.

Elastic $\stackrel{\rightarrow}{p}\stackrel{\rightarrow}{p}$ scattering may occur 
with one of the four polarization options ($S$,  $N$,  
$L$ or no polarization) for both the two protons in the entrance and 
exit channel. Basic symmetry and conservation principles of the strong interaction 
impose constraints such that only 25 out of 256 possible polarization observables can be 
linearly independent. Experimental data on at least nine of them at the same beam energy 
and scattering angle allow to determine the helicity amplitudes by a $\chi^2$-minimization, 
with the exception of an unobservable, global phase. Actually more than 9 observables are used 
for such a direct reconstruction, because not all of them are linearly independent and the 
impact of their uncertainties is minimized.

\begin{figure}
  \psfrag{theta}{\hspace{-1.5cm}$\theta_{c.m.}~(deg)$}
  \psfrag{direkte}{direct reconstruction}
  \psfrag{said}{\sc Said Fa00}
  \psfrag{phi1}{$|\phi_1|$}
  \psfrag{ang1}{$\alpha_1$}
  \psfrag{phi2}{$|\phi_2|$}
  \psfrag{ang2}{$\alpha_2$}
  \psfrag{phi3}{$|\phi_3|$}
  \psfrag{ang3}{$\alpha_3$}
  \psfrag{phi4}{$|\phi_4|$}
  \psfrag{ang4}{$\alpha_4$}
  \psfrag{phi5}{$|\phi_5|$}
  \psfrag{ang5}{$\alpha_5$}
  \psfrag{1}{\tiny$\pi$}
  \psfrag{0.5}{\tiny$\pi/2$}
  \psfrag{-0.5}{\hspace{-.1cm}\tiny$-\pi/2$}
  \psfrag{-1}{\hspace{-.1cm}\tiny$-\pi$}
  \centerline{\includegraphics[width=8.8cm]{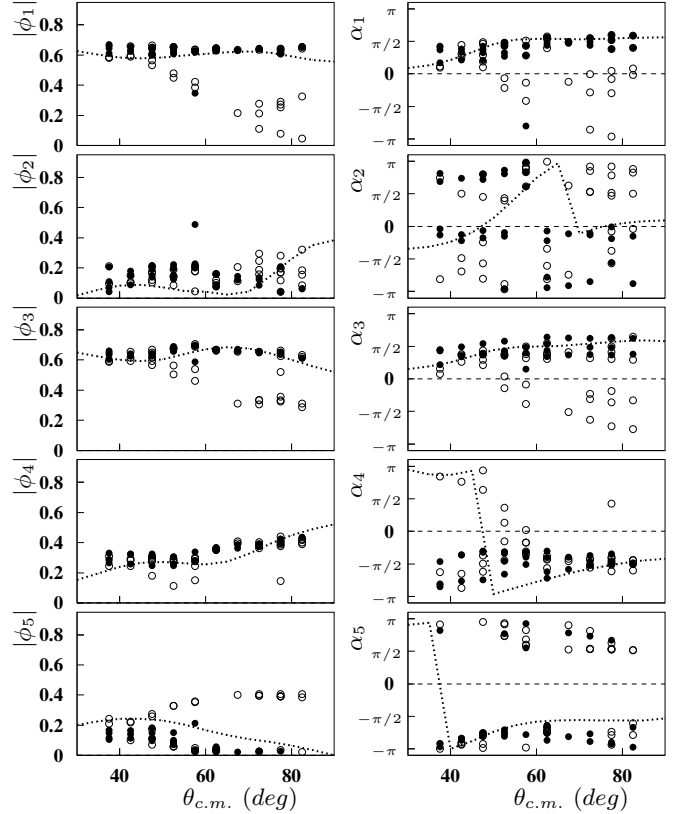}}
  \caption{Direct reconstruction of scattering amplitudes at
    $T_{lab}=$ 1.8~GeV
    with (filled circles) and without (open circles) inclusion of the
    new EDDA data. Different solutions at one angle reflect the
    minimum $\chi^2$ ambiguities.
    The solutions at different angles are independent.
    Phases are defined in the interval $-\pi\le\alpha_i<\pi$.
    This gives rise to spurious discontinuities e.g. for 
    $\alpha_4$ and $\alpha_5$ at $\pm \pi$.
    The dotted lines give the PSA solution FA00.
    }
  \label{fig:direkte}
\end{figure}

The EDDA data have been added to the world data base. For narrow energy intervals  
at 1.3, 1.6, 1.8, 2.1, and 2.4~GeV there are now 16 or more observables available that 
allow a direct reconstruction over a wide angular range. Results for 2.1~GeV were reported 
in \cite{EDD03}; here we emphasize the reconstructions at 1.8~GeV with up to 21 observables from 
\cite{LEH87,BAL99,ALL00,PER88,FON89,ALL98a,BAL99a,ALL99a,KOB94,LAC89b,LAC89a,LAC89c,LAC89d,LEH88,PER87}
, among them 11 with double and 8 with triple polarization information. 
Since the direct reconstruction is not a global phase shift analysis, it can easily lead 
to several solutions that describe the data equally well. The $\chi^2$ also is a measure of how 
the new data fits into the existing data base. Similar to the results for 2.1~GeV \cite{EDD03} and the 
findings in \cite{BYS98} we have obtained between one and four solutions in most cases.
Best results are achieved at lower energies (1.3 - 1.8~GeV).

\begin{figure}
  \psfrag{theta}{\hspace{-1.5cm}$\theta_{c.m.}~(deg)$}
  \psfrag{direkte}{direct reconstruction}
  \psfrag{said}{\sc Said Fa00}
  \psfrag{phi1}{$|\phi_1|$}
  \psfrag{ang1}{$\alpha_1$}
  \psfrag{phi2}{$|\phi_2|$}
  \psfrag{ang2}{$\alpha_2$}
  \psfrag{phi3}{$|\phi_3|$}
  \psfrag{ang3}{$\alpha_3$}
  \psfrag{phi4}{$|\phi_4|$}
  \psfrag{ang4}{$\alpha_4$}
  \psfrag{phi5}{$|\phi_5|$}
  \psfrag{ang5}{$\alpha_5$}
  \psfrag{1}{\tiny$\pi$}
  \psfrag{0.5}{\tiny$\pi/2$}
  \psfrag{-0.5}{\hspace{-.1cm}\tiny$-\pi/2$}
  \psfrag{-1}{\hspace{-.1cm}\tiny$-\pi$}
  \centerline{\includegraphics[width=8.8cm]{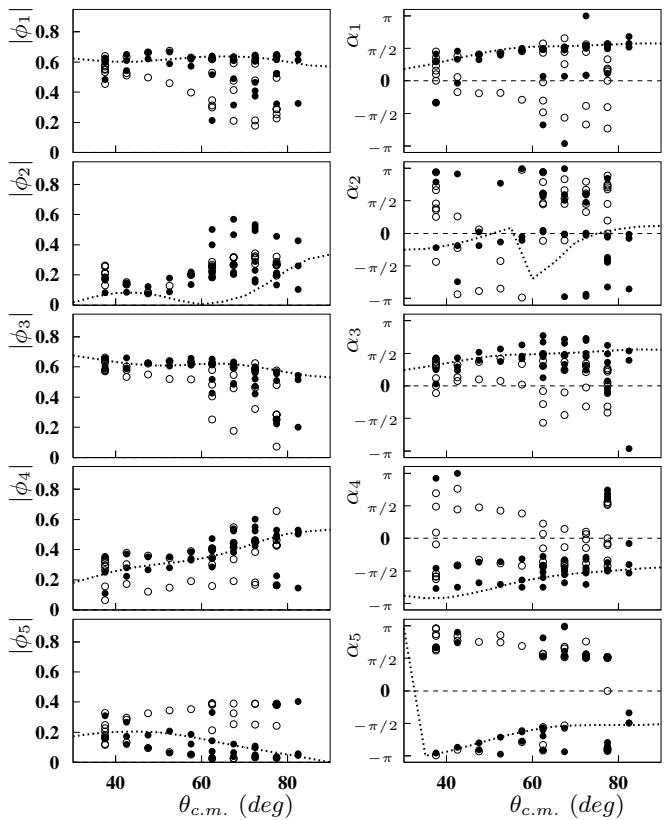}}
  \caption{DRSA at $T_{lab}=$ 1.6~GeV.
    The symbols have the same meaning as in Fig.\ \ref{fig:direkte}.
    }
  \label{fig:direkte2}
\end{figure}

Figure \ref{fig:direkte} shows the scattering amplitudes for 1.8~GeV in
terms of absolute values and directions in the complex plane 
\begin{equation}
\phi_k=|\phi_k|e^{i\alpha_k}
\label{eq24}
\end{equation} 
in comparison to the {\sc Said} phase shift solution FA00.
Solid (open) symbols denote DRSA solutions with (without) inclusion of
this work.
For normalization, the amplitudes are divided by
$(|\phi_1|^2+|\phi_2|^2+|\phi_3|^2+|\phi_4|^2+4|\phi_5|^2)^{\frac{1}{2}}=\sqrt{2I_0}$. 
Our new data do not increase the $\chi^2$ values of the reconstruction. 
This indicates that they - and in particular $A_{SS}$ in spite of the
prominent deviation of the PSA solutions - are compatible with the
existing experiments and provide additional constraints on phase
shifts and scattering amplitudes.
Please note that the inclusion of the correlation coefficients of this
work tends to concentrate the DRSA solutions on one of the two
branches visible without them.
The moduli $|\phi_i|$ of all amplitudes on the preferred branches are
well described through the PSA solution FA00 though differences exist
in detail.

The single spin flip amplitude $\phi_5$ is generally weak.
This, together with the phase differences
$|\alpha_5-\alpha_{1,3,4}|\approx\frac{\pi}{2}$, corresponds according 
to eq.(\ref{eq23}) to the small values found for $A_{SL}$.
Furthermore our DRSA yields that
$|\alpha_1-\alpha_2|\approx\frac{\pi}{2}$ or $\frac{3\pi}{2}$ implying
that  $Re(\phi_1\phi_2^{\star})\approx 0$.
From eqs.\ (\ref{eq21}) and (\ref{eq22}) follows then, that $A_{NN}$
and $A_{SS}$ are dominated by the bilinear product  $Re (\phi_3
\phi_4^{\star})$ of the amplitudes for no and double spin flip thus
preserving an initial antiparallel spin configuration, cf.\ eq.\ (\ref{eqn:helicity}).
The experimental result of $A_{SS} \approx -A_{NN}$ is but another
indication for an almost vanishing $|\phi_5|^2$. These findings confirm the 
results obtained in \cite{EDD03} at $T_{lab}$ = 2.1~GeV:
the single spin flip amplitude $\phi_5$, mainly driven by spin-orbit
forces \cite{CON94} is small at these energies and the amplitudes
$\phi_3$ with no and $\phi_4$ with double spin flip prevail.

In contrast to this result is the phase difference  $|\alpha_1-\alpha_2|$
of the solution {\sc Said} FA00 in Fig.\ \ref{fig:direkte} smaller
than $\frac{\pi}{2}$ such that $Re(\phi_1\phi_2^{\star})$ contributes
and the resulting $A_{SS}$ exceeds our experimental result
considerably.

Similiar conclusions can be drawn from the DRSA at 1.6~GeV (Fig.\
\ref{fig:direkte2}) such that a consistent picture emerges for
1.6 - 2.1~GeV.
At 2.4~GeV the data base is less restrictive and permits a
larger number of solutions.
This is mostly due to the scarce data of triple polarized observables, 
which have been measured at four angles only.

\section{Summary}
\label{sec:6}
The recirculating COSY beam has been used to study the elastic
scattering of polarized protons on a polarized atomic hydrogen target
{\it during acceleration} at beam energies between 0.45 and 2.50 GeV.
The highly granulated EDDA detector covered the angular range
30$^{\circ}\le\theta_{c.m.}\le$ 90$^{\circ}$; it was not only used to
identify elastic scattering events, but served also as internal
polarimeter monitoring the beam polarization during acceleration.
In addition data were taken at ten fixed energies between 0.77 and
2.44 GeV.
Absolute beam polarizations were obtained with reference to the
analyzing power excitation functions $A_N(T_{lab}, \theta_{c.m.})$ derived
previously with the same setup using an unpolarized beam and a
polarized target \cite{EDD05}.

Excitation functions of the spin correlation coefficients $A_{NN}$,
$A_{SS}$, and $A_{SL}$ have been determined over the whole energy and
angular range.
Those for $A_{NN}$ and $A_{SL}$ are mostly in reasonable agreement
with previous data and PSA solutions. 
For $A_{SS}$, however, previous data for PSA analyses were restricted to energies $T_{lab}
\le$ 0.79 GeV, and the PSA solutions based on them are the more at
variance with our data (and with one another) the more $T_{lab}$ exceeds
this energy.
We conclude that the previous world data base was insufficient to
allow an extrapolation of PSA solutions into regions not represented
in the data set.
The data can  be accessed via Ref. \cite{WWW04}. 

The direct reconstructions of scattering amplitudes for selected
energies discussed in Sec.\ \ref{sec:53} 
indicate, that the addition of our excitation functions for $A_{NN}$, $A_{SS}$, and $A_{SL}$ to the world 
data set will reduce the ambiguities in PSA solutions and thus improve their reliability and predictive 
power. 

\acknowledgments{
The EDDA collaboration thanks the COSY operating team for 
excellent, continuous beam support. Helpful discussions with R.A.\ Arndt and R.\ Machleidt 
are very much appreciated. This work was supported by the BMBF, contracts 06BN664I(6), 
06HH952, and 06HH152, and by the Forschungszentrum J\"ulich under FFE contracts 41126803, 41126903,
and 41520732.}

\begin{longtable}{|c|ccc|ccc|}
  \caption{Spin correlation parameters $A_{NN}$, $A_{SS}$ and $A_{SL}$
    for the 10 flattop energies
    \label{table-spincorr}} \\ \hline

& \multicolumn{3}{c|}{$p_{lab}=$1430 MeV/c}
& \multicolumn{3}{c|}{$p_{lab}=$1950 MeV/c} \\
& \multicolumn{3}{c|}{$\Delta_{norm}/norm=$ 2.7\%}
& \multicolumn{3}{c|}{$\Delta_{norm}/norm=$ 4.3\%} \\
$\theta_{c.m.}$&\multicolumn{3}{c|}{$A_{NN}$}&\multicolumn{3}{c|}{$A_{NN}$}\\ \hline 
32.5$^\circ$ & 0.404 & $\pm 0.056$ & $\pm 0.007$ & 0.485 & $\pm 0.057$ & $\pm 0.008$\\ \hline 
37.5$^\circ$ & 0.504 & $\pm 0.030$ & $\pm 0.006$ & 0.536 & $\pm 0.050$ & $\pm 0.007$\\ \hline 
42.5$^\circ$ & 0.512 & $\pm 0.023$ & $\pm 0.005$ & 0.545 & $\pm 0.045$ & $\pm 0.006$\\ \hline 
47.5$^\circ$ & 0.521 & $\pm 0.023$ & $\pm 0.005$ & 0.482 & $\pm 0.050$ & $\pm 0.006$\\ \hline 
52.5$^\circ$ & 0.527 & $\pm 0.024$ & $\pm 0.004$ & 0.590 & $\pm 0.057$ & $\pm 0.007$\\ \hline 
57.5$^\circ$ & 0.616 & $\pm 0.025$ & $\pm 0.004$ & 0.579 & $\pm 0.065$ & $\pm 0.009$\\ \hline 
62.5$^\circ$ & 0.562 & $\pm 0.027$ & $\pm 0.004$ & 0.486 & $\pm 0.074$ & $\pm 0.012$\\ \hline 
67.5$^\circ$ & 0.598 & $\pm 0.029$ & $\pm 0.004$ & 0.722 & $\pm 0.078$ & $\pm 0.016$\\ \hline 
72.5$^\circ$ & 0.618 & $\pm 0.031$ & $\pm 0.005$ & 0.446 & $\pm 0.090$ & $\pm 0.022$\\ \hline 
77.5$^\circ$ & 0.600 & $\pm 0.034$ & $\pm 0.006$ & 0.522 & $\pm 0.093$ & $\pm 0.029$\\ \hline 
82.5$^\circ$ & 0.639 & $\pm 0.035$ & $\pm 0.007$ & 0.497 & $\pm 0.098$ & $\pm 0.037$\\ \hline 
87.5$^\circ$ & 0.720 & $\pm 0.036$ & $\pm 0.009$ & 0.644 & $\pm 0.100$ & $\pm 0.047$\\ \hline 
\hline
& \multicolumn{3}{c|}{$p_{lab}=$2096 MeV/c}
& \multicolumn{3}{c|}{$p_{lab}=$2300 MeV/c} \\
& \multicolumn{3}{c|}{$\Delta_{norm}/norm=$ 2.9\%}
& \multicolumn{3}{c|}{$\Delta_{norm}/norm=$ 3.1\%} \\
$\theta_{c.m.}$&\multicolumn{3}{c|}{$A_{NN}$}&\multicolumn{3}{c|}{$A_{NN}$}\\ \hline 
32.5$^\circ$ & 0.365 & $\pm 0.014$ & $\pm 0.003$ & 0.347 & $\pm 0.013$ & $\pm 0.002$\\ \hline 
37.5$^\circ$ & 0.404 & $\pm 0.012$ & $\pm 0.003$ & 0.388 & $\pm 0.012$ & $\pm 0.002$\\ \hline 
42.5$^\circ$ & 0.467 & $\pm 0.013$ & $\pm 0.004$ & 0.416 & $\pm 0.014$ & $\pm 0.003$\\ \hline 
47.5$^\circ$ & 0.446 & $\pm 0.016$ & $\pm 0.004$ & 0.432 & $\pm 0.017$ & $\pm 0.003$\\ \hline 
52.5$^\circ$ & 0.463 & $\pm 0.018$ & $\pm 0.004$ & 0.382 & $\pm 0.019$ & $\pm 0.004$\\ \hline 
57.5$^\circ$ & 0.467 & $\pm 0.022$ & $\pm 0.005$ & 0.385 & $\pm 0.023$ & $\pm 0.005$\\ \hline 
62.5$^\circ$ & 0.522 & $\pm 0.024$ & $\pm 0.005$ & 0.364 & $\pm 0.025$ & $\pm 0.007$\\ \hline 
67.5$^\circ$ & 0.457 & $\pm 0.026$ & $\pm 0.006$ & 0.403 & $\pm 0.028$ & $\pm 0.008$\\ \hline 
72.5$^\circ$ & 0.438 & $\pm 0.029$ & $\pm 0.007$ & 0.402 & $\pm 0.030$ & $\pm 0.010$\\ \hline 
77.5$^\circ$ & 0.515 & $\pm 0.030$ & $\pm 0.008$ & 0.492 & $\pm 0.031$ & $\pm 0.012$\\ \hline 
82.5$^\circ$ & 0.463 & $\pm 0.032$ & $\pm 0.009$ & 0.560 & $\pm 0.031$ & $\pm 0.015$\\ \hline 
87.5$^\circ$ & 0.520 & $\pm 0.031$ & $\pm 0.010$ & 0.576 & $\pm 0.032$ & $\pm 0.017$\\ \hline 
\hline
& \multicolumn{3}{c|}{$p_{lab}=$2572 MeV/c}
& \multicolumn{3}{c|}{$p_{lab}=$2720 MeV/c} \\
& \multicolumn{3}{c|}{$\Delta_{norm}/norm=$ 3.4\%}
& \multicolumn{3}{c|}{$\Delta_{norm}/norm=$ 4.2\%} \\
$\theta_{c.m.}$&\multicolumn{3}{c|}{$A_{NN}$}&\multicolumn{3}{c|}{$A_{NN}$}\\ \hline 
32.5$^\circ$ & 0.312 & $\pm 0.012$ & $\pm 0.002$ & 0.264 & $\pm 0.013$ & $\pm 0.003$\\ \hline 
37.5$^\circ$ & 0.317 & $\pm 0.012$ & $\pm 0.002$ & 0.290 & $\pm 0.012$ & $\pm 0.003$\\ \hline 
42.5$^\circ$ & 0.305 & $\pm 0.014$ & $\pm 0.002$ & 0.276 & $\pm 0.016$ & $\pm 0.003$\\ \hline 
47.5$^\circ$ & 0.294 & $\pm 0.018$ & $\pm 0.002$ & 0.248 & $\pm 0.019$ & $\pm 0.004$\\ \hline 
52.5$^\circ$ & 0.296 & $\pm 0.021$ & $\pm 0.003$ & 0.229 & $\pm 0.023$ & $\pm 0.004$\\ \hline 
57.5$^\circ$ & 0.292 & $\pm 0.025$ & $\pm 0.004$ & 0.185 & $\pm 0.027$ & $\pm 0.005$\\ \hline 
62.5$^\circ$ & 0.273 & $\pm 0.026$ & $\pm 0.005$ & 0.219 & $\pm 0.028$ & $\pm 0.006$\\ \hline 
67.5$^\circ$ & 0.322 & $\pm 0.030$ & $\pm 0.006$ & 0.316 & $\pm 0.031$ & $\pm 0.008$\\ \hline 
72.5$^\circ$ & 0.347 & $\pm 0.030$ & $\pm 0.008$ & 0.350 & $\pm 0.032$ & $\pm 0.010$\\ \hline 
77.5$^\circ$ & 0.419 & $\pm 0.031$ & $\pm 0.010$ & 0.414 & $\pm 0.032$ & $\pm 0.012$\\ \hline 
82.5$^\circ$ & 0.514 & $\pm 0.033$ & $\pm 0.013$ & 0.467 & $\pm 0.035$ & $\pm 0.015$\\ \hline 
87.5$^\circ$ & 0.523 & $\pm 0.032$ & $\pm 0.016$ & 0.502 & $\pm 0.034$ & $\pm 0.019$\\ \hline 
\hline
& \multicolumn{3}{c|}{$p_{lab}=$2900 MeV/c}
& \multicolumn{3}{c|}{$p_{lab}=$3100 MeV/c} \\
& \multicolumn{3}{c|}{$\Delta_{norm}/norm=$ 4.2\%}
& \multicolumn{3}{c|}{$\Delta_{norm}/norm=$ 4.8\%} \\
$\theta_{c.m.}$&\multicolumn{3}{c|}{$A_{NN}$}&\multicolumn{3}{c|}{$A_{NN}$}\\ \hline 
32.5$^\circ$ & 0.267 & $\pm 0.018$ & $\pm 0.005$ & 0.248 & $\pm 0.022$ & $\pm 0.003$\\ \hline 
37.5$^\circ$ & 0.227 & $\pm 0.018$ & $\pm 0.005$ & 0.202 & $\pm 0.022$ & $\pm 0.004$\\ \hline 
42.5$^\circ$ & 0.227 & $\pm 0.023$ & $\pm 0.005$ & 0.208 & $\pm 0.028$ & $\pm 0.005$\\ \hline 
47.5$^\circ$ & 0.230 & $\pm 0.029$ & $\pm 0.005$ & 0.212 & $\pm 0.034$ & $\pm 0.006$\\ \hline 
52.5$^\circ$ & 0.277 & $\pm 0.036$ & $\pm 0.006$ & 0.276 & $\pm 0.042$ & $\pm 0.008$\\ \hline 
57.5$^\circ$ & 0.206 & $\pm 0.040$ & $\pm 0.008$ & 0.219 & $\pm 0.047$ & $\pm 0.010$\\ \hline 
62.5$^\circ$ & 0.285 & $\pm 0.042$ & $\pm 0.009$ & 0.183 & $\pm 0.052$ & $\pm 0.012$\\ \hline 
67.5$^\circ$ & 0.326 & $\pm 0.046$ & $\pm 0.012$ & 0.289 & $\pm 0.055$ & $\pm 0.014$\\ \hline 
72.5$^\circ$ & 0.274 & $\pm 0.049$ & $\pm 0.015$ & 0.316 & $\pm 0.058$ & $\pm 0.016$\\ \hline 
77.5$^\circ$ & 0.371 & $\pm 0.051$ & $\pm 0.019$ & 0.227 & $\pm 0.062$ & $\pm 0.019$\\ \hline 
82.5$^\circ$ & 0.478 & $\pm 0.053$ & $\pm 0.024$ & 0.568 & $\pm 0.064$ & $\pm 0.022$\\ \hline 
87.5$^\circ$ & 0.441 & $\pm 0.055$ & $\pm 0.030$ & 0.418 & $\pm 0.064$ & $\pm 0.025$\\ \hline 
\hline
& \multicolumn{3}{c|}{$p_{lab}=$3180 MeV/c}
& \multicolumn{3}{c|}{$p_{lab}=$3300 MeV/c} \\
& \multicolumn{3}{c|}{$\Delta_{norm}/norm=$ 4.4\%}
& \multicolumn{3}{c|}{$\Delta_{norm}/norm=$ 3.3\%} \\
$\theta_{c.m.}$&\multicolumn{3}{c|}{$A_{NN}$}&\multicolumn{3}{c|}{$A_{NN}$}\\ \hline 
32.5$^\circ$ & 0.264 & $\pm 0.016$ & $\pm 0.007$ & 0.157 & $\pm 0.017$ & $\pm 0.005$\\ \hline 
37.5$^\circ$ & 0.266 & $\pm 0.017$ & $\pm 0.008$ & 0.163 & $\pm 0.017$ & $\pm 0.007$\\ \hline 
42.5$^\circ$ & 0.233 & $\pm 0.022$ & $\pm 0.009$ & 0.205 & $\pm 0.022$ & $\pm 0.008$\\ \hline 
47.5$^\circ$ & 0.235 & $\pm 0.028$ & $\pm 0.010$ & 0.176 & $\pm 0.028$ & $\pm 0.010$\\ \hline 
52.5$^\circ$ & 0.172 & $\pm 0.034$ & $\pm 0.011$ & 0.161 & $\pm 0.034$ & $\pm 0.012$\\ \hline 
57.5$^\circ$ & 0.307 & $\pm 0.036$ & $\pm 0.013$ & 0.215 & $\pm 0.036$ & $\pm 0.015$\\ \hline 
62.5$^\circ$ & 0.296 & $\pm 0.040$ & $\pm 0.015$ & 0.300 & $\pm 0.041$ & $\pm 0.018$\\ \hline 
67.5$^\circ$ & 0.386 & $\pm 0.044$ & $\pm 0.017$ & 0.241 & $\pm 0.045$ & $\pm 0.022$\\ \hline 
72.5$^\circ$ & 0.373 & $\pm 0.046$ & $\pm 0.019$ & 0.290 & $\pm 0.047$ & $\pm 0.026$\\ \hline 
77.5$^\circ$ & 0.375 & $\pm 0.047$ & $\pm 0.021$ & 0.334 & $\pm 0.051$ & $\pm 0.031$\\ \hline 
82.5$^\circ$ & 0.423 & $\pm 0.051$ & $\pm 0.024$ & 0.298 & $\pm 0.053$ & $\pm 0.036$\\ \hline 
87.5$^\circ$ & 0.431 & $\pm 0.051$ & $\pm 0.027$ & 0.330 & $\pm 0.057$ & $\pm 0.041$\\ \hline 
\hline
& \multicolumn{3}{c|}{$p_{lab}=$1430 MeV/c}
& \multicolumn{3}{c|}{$p_{lab}=$1950 MeV/c} \\
& \multicolumn{3}{c|}{$\Delta_{norm}/norm=$ 2.7\%}
& \multicolumn{3}{c|}{$\Delta_{norm}/norm=$ 4.3\%} \\
$\theta_{c.m.}$&\multicolumn{3}{c|}{$A_{SS}$}&\multicolumn{3}{c|}{$A_{SS}$}\\ \hline 
32.5$^\circ$ & -0.323 & $\pm 0.056$ & $\pm 0.011$ & -0.261 & $\pm 0.057$ & $\pm 0.010$\\ \hline 
37.5$^\circ$ & -0.516 & $\pm 0.031$ & $\pm 0.009$ & -0.425 & $\pm 0.051$ & $\pm 0.011$\\ \hline 
42.5$^\circ$ & -0.492 & $\pm 0.028$ & $\pm 0.008$ & -0.454 & $\pm 0.056$ & $\pm 0.012$\\ \hline 
47.5$^\circ$ & -0.487 & $\pm 0.029$ & $\pm 0.007$ & -0.346 & $\pm 0.064$ & $\pm 0.013$\\ \hline 
52.5$^\circ$ & -0.426 & $\pm 0.030$ & $\pm 0.006$ & -0.399 & $\pm 0.072$ & $\pm 0.014$\\ \hline 
57.5$^\circ$ & -0.548 & $\pm 0.032$ & $\pm 0.005$ & -0.435 & $\pm 0.082$ & $\pm 0.015$\\ \hline 
62.5$^\circ$ & -0.513 & $\pm 0.034$ & $\pm 0.004$ & -0.318 & $\pm 0.093$ & $\pm 0.015$\\ \hline 
67.5$^\circ$ & -0.494 & $\pm 0.037$ & $\pm 0.004$ & -0.540 & $\pm 0.095$ & $\pm 0.016$\\ \hline 
72.5$^\circ$ & -0.571 & $\pm 0.038$ & $\pm 0.003$ & -0.377 & $\pm 0.111$ & $\pm 0.017$\\ \hline 
77.5$^\circ$ & -0.533 & $\pm 0.042$ & $\pm 0.004$ & -0.725 & $\pm 0.116$ & $\pm 0.018$\\ \hline 
82.5$^\circ$ & -0.523 & $\pm 0.044$ & $\pm 0.004$ & -0.383 & $\pm 0.120$ & $\pm 0.019$\\ \hline 
87.5$^\circ$ & -0.482 & $\pm 0.044$ & $\pm 0.005$ & -0.757 & $\pm 0.123$ & $\pm 0.019$\\ \hline 
\hline
& \multicolumn{3}{c|}{$p_{lab}=$2096 MeV/c}
& \multicolumn{3}{c|}{$p_{lab}=$2300 MeV/c} \\
& \multicolumn{3}{c|}{$\Delta_{norm}/norm=$ 2.9\%}
& \multicolumn{3}{c|}{$\Delta_{norm}/norm=$ 3.1\%} \\
$\theta_{c.m.}$&\multicolumn{3}{c|}{$A_{SS}$}&\multicolumn{3}{c|}{$A_{SS}$}\\ \hline 
32.5$^\circ$ & -0.341 & $\pm 0.014$ & $\pm 0.001$ & -0.292 & $\pm 0.013$ & $\pm 0.002$\\ \hline 
37.5$^\circ$ & -0.352 & $\pm 0.015$ & $\pm 0.002$ & -0.278 & $\pm 0.015$ & $\pm 0.003$\\ \hline 
42.5$^\circ$ & -0.367 & $\pm 0.017$ & $\pm 0.003$ & -0.297 & $\pm 0.018$ & $\pm 0.004$\\ \hline 
47.5$^\circ$ & -0.349 & $\pm 0.020$ & $\pm 0.005$ & -0.292 & $\pm 0.021$ & $\pm 0.006$\\ \hline 
52.5$^\circ$ & -0.385 & $\pm 0.023$ & $\pm 0.006$ & -0.252 & $\pm 0.025$ & $\pm 0.007$\\ \hline 
57.5$^\circ$ & -0.330 & $\pm 0.027$ & $\pm 0.007$ & -0.341 & $\pm 0.030$ & $\pm 0.010$\\ \hline 
62.5$^\circ$ & -0.369 & $\pm 0.030$ & $\pm 0.009$ & -0.292 & $\pm 0.032$ & $\pm 0.012$\\ \hline 
67.5$^\circ$ & -0.367 & $\pm 0.033$ & $\pm 0.010$ & -0.421 & $\pm 0.035$ & $\pm 0.015$\\ \hline 
72.5$^\circ$ & -0.398 & $\pm 0.036$ & $\pm 0.012$ & -0.421 & $\pm 0.037$ & $\pm 0.019$\\ \hline 
77.5$^\circ$ & -0.464 & $\pm 0.038$ & $\pm 0.013$ & -0.562 & $\pm 0.039$ & $\pm 0.022$\\ \hline 
82.5$^\circ$ & -0.477 & $\pm 0.040$ & $\pm 0.015$ & -0.594 & $\pm 0.039$ & $\pm 0.026$\\ \hline 
87.5$^\circ$ & -0.484 & $\pm 0.039$ & $\pm 0.017$ & -0.643 & $\pm 0.041$ & $\pm 0.031$\\ \hline 
\hline
& \multicolumn{3}{c|}{$p_{lab}=$2572 MeV/c}
& \multicolumn{3}{c|}{$p_{lab}=$2720 MeV/c} \\
& \multicolumn{3}{c|}{$\Delta_{norm}/norm=$ 3.4\%}
& \multicolumn{3}{c|}{$\Delta_{norm}/norm=$ 4.2\%} \\
$\theta_{c.m.}$&\multicolumn{3}{c|}{$A_{SS}$}&\multicolumn{3}{c|}{$A_{SS}$}\\ \hline 
32.5$^\circ$ & -0.229 & $\pm 0.013$ & $\pm 0.002$ & -0.219 & $\pm 0.013$ & $\pm 0.001$\\ \hline 
37.5$^\circ$ & -0.227 & $\pm 0.015$ & $\pm 0.002$ & -0.238 & $\pm 0.016$ & $\pm 0.004$\\ \hline 
42.5$^\circ$ & -0.257 & $\pm 0.019$ & $\pm 0.003$ & -0.244 & $\pm 0.020$ & $\pm 0.006$\\ \hline 
47.5$^\circ$ & -0.221 & $\pm 0.023$ & $\pm 0.004$ & -0.274 & $\pm 0.024$ & $\pm 0.009$\\ \hline 
52.5$^\circ$ & -0.298 & $\pm 0.027$ & $\pm 0.005$ & -0.262 & $\pm 0.030$ & $\pm 0.012$\\ \hline 
57.5$^\circ$ & -0.351 & $\pm 0.032$ & $\pm 0.007$ & -0.281 & $\pm 0.033$ & $\pm 0.014$\\ \hline 
62.5$^\circ$ & -0.377 & $\pm 0.033$ & $\pm 0.009$ & -0.297 & $\pm 0.035$ & $\pm 0.017$\\ \hline 
67.5$^\circ$ & -0.410 & $\pm 0.037$ & $\pm 0.011$ & -0.413 & $\pm 0.038$ & $\pm 0.019$\\ \hline 
72.5$^\circ$ & -0.457 & $\pm 0.038$ & $\pm 0.013$ & -0.421 & $\pm 0.039$ & $\pm 0.020$\\ \hline 
77.5$^\circ$ & -0.459 & $\pm 0.040$ & $\pm 0.016$ & -0.471 & $\pm 0.042$ & $\pm 0.022$\\ \hline 
82.5$^\circ$ & -0.509 & $\pm 0.041$ & $\pm 0.019$ & -0.430 & $\pm 0.043$ & $\pm 0.022$\\ \hline 
87.5$^\circ$ & -0.505 & $\pm 0.040$ & $\pm 0.022$ & -0.440 & $\pm 0.042$ & $\pm 0.022$\\ \hline 
\hline
& \multicolumn{3}{c|}{$p_{lab}=$2900 MeV/c}
& \multicolumn{3}{c|}{$p_{lab}=$3100 MeV/c} \\
& \multicolumn{3}{c|}{$\Delta_{norm}/norm=$ 4.2\%}
& \multicolumn{3}{c|}{$\Delta_{norm}/norm=$ 4.8\%} \\
$\theta_{c.m.}$&\multicolumn{3}{c|}{$A_{SS}$}&\multicolumn{3}{c|}{$A_{SS}$}\\ \hline 
32.5$^\circ$ & -0.196 & $\pm 0.018$ & $\pm 0.000$ & -0.222 & $\pm 0.022$ & $\pm 0.004$\\ \hline 
37.5$^\circ$ & -0.194 & $\pm 0.023$ & $\pm 0.005$ & -0.189 & $\pm 0.029$ & $\pm 0.005$\\ \hline 
42.5$^\circ$ & -0.239 & $\pm 0.030$ & $\pm 0.009$ & -0.249 & $\pm 0.036$ & $\pm 0.005$\\ \hline 
47.5$^\circ$ & -0.258 & $\pm 0.037$ & $\pm 0.013$ & -0.303 & $\pm 0.044$ & $\pm 0.006$\\ \hline 
52.5$^\circ$ & -0.308 & $\pm 0.045$ & $\pm 0.017$ & -0.395 & $\pm 0.053$ & $\pm 0.007$\\ \hline 
57.5$^\circ$ & -0.306 & $\pm 0.050$ & $\pm 0.020$ & -0.321 & $\pm 0.058$ & $\pm 0.009$\\ \hline 
62.5$^\circ$ & -0.308 & $\pm 0.052$ & $\pm 0.022$ & -0.357 & $\pm 0.064$ & $\pm 0.010$\\ \hline 
67.5$^\circ$ & -0.508 & $\pm 0.056$ & $\pm 0.023$ & -0.464 & $\pm 0.068$ & $\pm 0.011$\\ \hline 
72.5$^\circ$ & -0.419 & $\pm 0.060$ & $\pm 0.024$ & -0.432 & $\pm 0.070$ & $\pm 0.013$\\ \hline 
77.5$^\circ$ & -0.458 & $\pm 0.063$ & $\pm 0.023$ & -0.311 & $\pm 0.074$ & $\pm 0.015$\\ \hline 
82.5$^\circ$ & -0.353 & $\pm 0.063$ & $\pm 0.020$ & -0.559 & $\pm 0.076$ & $\pm 0.016$\\ \hline 
87.5$^\circ$ & -0.459 & $\pm 0.065$ & $\pm 0.017$ & -0.593 & $\pm 0.078$ & $\pm 0.018$\\ \hline 
\hline
& \multicolumn{3}{c|}{$p_{lab}=$3180 MeV/c}
& \multicolumn{3}{c|}{$p_{lab}=$3300 MeV/c} \\
& \multicolumn{3}{c|}{$\Delta_{norm}/norm=$ 4.4\%}
& \multicolumn{3}{c|}{$\Delta_{norm}/norm=$ 3.3\%} \\
$\theta_{c.m.}$&\multicolumn{3}{c|}{$A_{SS}$}&\multicolumn{3}{c|}{$A_{SS}$}\\ \hline 
32.5$^\circ$ & -0.194 & $\pm 0.017$ & $\pm 0.006$ & -0.160 & $\pm 0.017$ & $\pm 0.005$\\ \hline
37.5$^\circ$ & -0.207 & $\pm 0.022$ & $\pm 0.007$ & -0.182 & $\pm 0.022$ & $\pm 0.008$\\ \hline 
42.5$^\circ$ & -0.288 & $\pm 0.029$ & $\pm 0.009$ & -0.249 & $\pm 0.029$ & $\pm 0.012$\\ \hline 
47.5$^\circ$ & -0.262 & $\pm 0.036$ & $\pm 0.012$ & -0.265 & $\pm 0.035$ & $\pm 0.015$\\ \hline 
52.5$^\circ$ & -0.274 & $\pm 0.044$ & $\pm 0.014$ & -0.218 & $\pm 0.042$ & $\pm 0.019$\\ \hline 
57.5$^\circ$ & -0.317 & $\pm 0.045$ & $\pm 0.017$ & -0.391 & $\pm 0.044$ & $\pm 0.022$\\ \hline 
62.5$^\circ$ & -0.443 & $\pm 0.050$ & $\pm 0.020$ & -0.422 & $\pm 0.049$ & $\pm 0.025$\\ \hline 
67.5$^\circ$ & -0.549 & $\pm 0.054$ & $\pm 0.024$ & -0.433 & $\pm 0.054$ & $\pm 0.027$\\ \hline 
72.5$^\circ$ & -0.442 & $\pm 0.058$ & $\pm 0.028$ & -0.382 & $\pm 0.056$ & $\pm 0.029$\\ \hline 
77.5$^\circ$ & -0.519 & $\pm 0.061$ & $\pm 0.032$ & -0.423 & $\pm 0.061$ & $\pm 0.030$\\ \hline 
82.5$^\circ$ & -0.531 & $\pm 0.064$ & $\pm 0.036$ & -0.473 & $\pm 0.063$ & $\pm 0.030$\\ \hline 
87.5$^\circ$ & -0.627 & $\pm 0.064$ & $\pm 0.041$ & -0.495 & $\pm 0.066$ & $\pm 0.030$\\ \hline 
\hline
& \multicolumn{3}{c|}{$p_{lab}=$1430 MeV/c}
& \multicolumn{3}{c|}{$p_{lab}=$1950 MeV/c} \\
& \multicolumn{3}{c|}{$\Delta_{norm}/norm=$ 2.7\%}
& \multicolumn{3}{c|}{$\Delta_{norm}/norm=$ 4.3\%} \\
$\theta_{c.m.}$&\multicolumn{3}{c|}{$A_{SL}$}&\multicolumn{3}{c|}{$A_{SL}$}\\ \hline 
32.5$^\circ$ & -0.001 & $\pm 0.053$ & $\pm 0.011$ & -0.083 & $\pm 0.055$ & $\pm 0.002$\\ \hline 
37.5$^\circ$ & -0.072 & $\pm 0.030$ & $\pm 0.009$ & -0.101 & $\pm 0.050$ & $\pm 0.005$\\ \hline 
42.5$^\circ$ & -0.018 & $\pm 0.026$ & $\pm 0.008$ & -0.064 & $\pm 0.052$ & $\pm 0.008$\\ \hline 
47.5$^\circ$ & -0.039 & $\pm 0.026$ & $\pm 0.007$ & -0.092 & $\pm 0.057$ & $\pm 0.011$\\ \hline 
52.5$^\circ$ & -0.030 & $\pm 0.027$ & $\pm 0.006$ & -0.030 & $\pm 0.065$ & $\pm 0.014$\\ \hline 
57.5$^\circ$ & -0.039 & $\pm 0.029$ & $\pm 0.006$ & -0.008 & $\pm 0.077$ & $\pm 0.017$\\ \hline 
62.5$^\circ$ &  0.014 & $\pm 0.031$ & $\pm 0.005$ & -0.006 & $\pm 0.086$ & $\pm 0.019$\\ \hline 
67.5$^\circ$ &  0.022 & $\pm 0.034$ & $\pm 0.005$ & -0.118 & $\pm 0.087$ & $\pm 0.020$\\ \hline 
72.5$^\circ$ & -0.043 & $\pm 0.035$ & $\pm 0.005$ &  0.047 & $\pm 0.100$ & $\pm 0.020$\\ \hline 
77.5$^\circ$ & -0.049 & $\pm 0.038$ & $\pm 0.006$ & -0.160 & $\pm 0.108$ & $\pm 0.020$\\ \hline 
82.5$^\circ$ &  0.077 & $\pm 0.040$ & $\pm 0.007$ &  0.038 & $\pm 0.109$ & $\pm 0.019$\\ \hline 
87.5$^\circ$ & -0.014 & $\pm 0.040$ & $\pm 0.009$ & -0.074 & $\pm 0.113$ & $\pm 0.018$\\ \hline 
\hline
& \multicolumn{3}{c|}{$p_{lab}=$2096 MeV/c}
& \multicolumn{3}{c|}{$p_{lab}=$2300 MeV/c} \\
& \multicolumn{3}{c|}{$\Delta_{norm}/norm=$ 2.9\%}
& \multicolumn{3}{c|}{$\Delta_{norm}/norm=$ 3.1\%} \\
$\theta_{c.m.}$&\multicolumn{3}{c|}{$A_{SL}$}&\multicolumn{3}{c|}{$A_{SL}$}\\ \hline 
32.5$^\circ$ & -0.038 & $\pm 0.014$ & $\pm 0.004$ & -0.042 & $\pm 0.013$ & $\pm 0.001$\\ \hline 
37.5$^\circ$ & -0.033 & $\pm 0.014$ & $\pm 0.004$ & -0.024 & $\pm 0.014$ & $\pm 0.002$\\ \hline 
42.5$^\circ$ & -0.017 & $\pm 0.016$ & $\pm 0.003$ &  0.004 & $\pm 0.016$ & $\pm 0.002$\\ \hline 
47.5$^\circ$ & -0.028 & $\pm 0.018$ & $\pm 0.003$ & -0.005 & $\pm 0.019$ & $\pm 0.003$\\ \hline 
52.5$^\circ$ &  0.018 & $\pm 0.021$ & $\pm 0.003$ &  0.018 & $\pm 0.023$ & $\pm 0.004$\\ \hline 
57.5$^\circ$ &  0.019 & $\pm 0.025$ & $\pm 0.004$ &  0.077 & $\pm 0.027$ & $\pm 0.005$\\ \hline 
62.5$^\circ$ &  0.008 & $\pm 0.027$ & $\pm 0.004$ &  0.017 & $\pm 0.029$ & $\pm 0.006$\\ \hline 
67.5$^\circ$ &  0.037 & $\pm 0.030$ & $\pm 0.004$ &  0.003 & $\pm 0.031$ & $\pm 0.007$\\ \hline 
72.5$^\circ$ &  0.007 & $\pm 0.033$ & $\pm 0.005$ & -0.029 & $\pm 0.034$ & $\pm 0.008$\\ \hline 
77.5$^\circ$ & -0.006 & $\pm 0.034$ & $\pm 0.005$ & -0.029 & $\pm 0.036$ & $\pm 0.009$\\ \hline 
82.5$^\circ$ &  0.026 & $\pm 0.036$ & $\pm 0.006$ &  0.043 & $\pm 0.036$ & $\pm 0.009$\\ \hline 
87.5$^\circ$ & -0.032 & $\pm 0.036$ & $\pm 0.007$ &  0.099 & $\pm 0.037$ & $\pm 0.010$\\ \hline 
\hline
& \multicolumn{3}{c|}{$p_{lab}=$2572 MeV/c}
& \multicolumn{3}{c|}{$p_{lab}=$2720 MeV/c} \\
& \multicolumn{3}{c|}{$\Delta_{norm}/norm=$ 3.4\%}
& \multicolumn{3}{c|}{$\Delta_{norm}/norm=$ 4.2\%} \\
$\theta_{c.m.}$&\multicolumn{3}{c|}{$A_{SL}$}&\multicolumn{3}{c|}{$A_{SL}$}\\ \hline 
32.5$^\circ$ & -0.019 & $\pm 0.012$ & $\pm 0.002$ & -0.035 & $\pm 0.012$ & $\pm 0.002$\\ \hline 
37.5$^\circ$ &  0.010 & $\pm 0.014$ & $\pm 0.002$ & -0.023 & $\pm 0.015$ & $\pm 0.002$\\ \hline 
42.5$^\circ$ & -0.024 & $\pm 0.017$ & $\pm 0.002$ &  0.014 & $\pm 0.018$ & $\pm 0.003$\\ \hline 
47.5$^\circ$ & -0.015 & $\pm 0.020$ & $\pm 0.003$ & -0.010 & $\pm 0.022$ & $\pm 0.003$\\ \hline 
52.5$^\circ$ &  0.007 & $\pm 0.025$ & $\pm 0.004$ &  0.013 & $\pm 0.027$ & $\pm 0.004$\\ \hline 
57.5$^\circ$ &  0.004 & $\pm 0.029$ & $\pm 0.004$ &  0.025 & $\pm 0.030$ & $\pm 0.005$\\ \hline 
62.5$^\circ$ &  0.009 & $\pm 0.030$ & $\pm 0.005$ & -0.042 & $\pm 0.032$ & $\pm 0.006$\\ \hline 
67.5$^\circ$ & -0.007 & $\pm 0.033$ & $\pm 0.006$ &  0.059 & $\pm 0.035$ & $\pm 0.008$\\ \hline 
72.5$^\circ$ & -0.059 & $\pm 0.034$ & $\pm 0.007$ & -0.047 & $\pm 0.036$ & $\pm 0.010$\\ \hline 
77.5$^\circ$ & -0.034 & $\pm 0.037$ & $\pm 0.009$ & -0.022 & $\pm 0.038$ & $\pm 0.012$\\ \hline 
82.5$^\circ$ & -0.022 & $\pm 0.037$ & $\pm 0.010$ & -0.022 & $\pm 0.039$ & $\pm 0.014$\\ \hline 
87.5$^\circ$ & -0.023 & $\pm 0.037$ & $\pm 0.012$ & -0.071 & $\pm 0.039$ & $\pm 0.017$\\ \hline 
\hline
& \multicolumn{3}{c|}{$p_{lab}=$2900 MeV/c}
& \multicolumn{3}{c|}{$p_{lab}=$3100 MeV/c} \\
& \multicolumn{3}{c|}{$\Delta_{norm}/norm=$ 4.2\%}
& \multicolumn{3}{c|}{$\Delta_{norm}/norm=$ 4.8\%} \\
$\theta_{c.m.}$&\multicolumn{3}{c|}{$A_{SL}$}&\multicolumn{3}{c|}{$A_{SL}$}\\ \hline 
32.5$^\circ$ & -0.026 & $\pm 0.017$ & $\pm 0.002$ & -0.009 & $\pm 0.021$ & $\pm 0.002$\\ \hline 
37.5$^\circ$ &  0.008 & $\pm 0.021$ & $\pm 0.003$ & -0.016 & $\pm 0.026$ & $\pm 0.002$\\ \hline 
42.5$^\circ$ & -0.041 & $\pm 0.027$ & $\pm 0.005$ & -0.022 & $\pm 0.032$ & $\pm 0.003$\\ \hline 
47.5$^\circ$ &  0.034 & $\pm 0.033$ & $\pm 0.007$ & -0.012 & $\pm 0.039$ & $\pm 0.004$\\ \hline 
52.5$^\circ$ &  0.060 & $\pm 0.041$ & $\pm 0.008$ & -0.022 & $\pm 0.047$ & $\pm 0.005$\\ \hline 
57.5$^\circ$ & -0.065 & $\pm 0.045$ & $\pm 0.010$ &  0.121 & $\pm 0.051$ & $\pm 0.006$\\ \hline 
62.5$^\circ$ & -0.038 & $\pm 0.047$ & $\pm 0.012$ &  0.031 & $\pm 0.057$ & $\pm 0.008$\\ \hline 
67.5$^\circ$ & -0.131 & $\pm 0.051$ & $\pm 0.013$ & -0.033 & $\pm 0.060$ & $\pm 0.011$\\ \hline 
72.5$^\circ$ & -0.002 & $\pm 0.054$ & $\pm 0.015$ &  0.068 & $\pm 0.064$ & $\pm 0.014$\\ \hline 
77.5$^\circ$ & -0.039 & $\pm 0.057$ & $\pm 0.016$ &  0.038 & $\pm 0.066$ & $\pm 0.018$\\ \hline 
82.5$^\circ$ & -0.124 & $\pm 0.058$ & $\pm 0.018$ & -0.073 & $\pm 0.068$ & $\pm 0.022$\\ \hline 
87.5$^\circ$ & -0.058 & $\pm 0.059$ & $\pm 0.019$ & -0.022 & $\pm 0.072$ & $\pm 0.027$\\ \hline 
\hline
& \multicolumn{3}{c|}{$p_{lab}=$3180 MeV/c}
& \multicolumn{3}{c|}{$p_{lab}=$3300 MeV/c} \\
& \multicolumn{3}{c|}{$\Delta_{norm}/norm=$ 4.4\%}
& \multicolumn{3}{c|}{$\Delta_{norm}/norm=$ 3.3\%} \\
$\theta_{c.m.}$&\multicolumn{3}{c|}{$A_{SL}$}&\multicolumn{3}{c|}{$A_{SL}$}\\ \hline 
32.5$^\circ$ & -0.008 & $\pm 0.016$ & $\pm 0.003$ & -0.013 & $\pm 0.016$ & $\pm 0.003$\\ \hline 
37.5$^\circ$ & -0.036 & $\pm 0.020$ & $\pm 0.005$ & -0.012 & $\pm 0.020$ & $\pm 0.005$\\ \hline 
42.5$^\circ$ &  0.017 & $\pm 0.026$ & $\pm 0.007$ & -0.010 & $\pm 0.026$ & $\pm 0.007$\\ \hline 
47.5$^\circ$ &  0.024 & $\pm 0.032$ & $\pm 0.009$ &  0.040 & $\pm 0.032$ & $\pm 0.009$\\ \hline 
52.5$^\circ$ & -0.023 & $\pm 0.040$ & $\pm 0.011$ & -0.073 & $\pm 0.037$ & $\pm 0.012$\\ \hline 
57.5$^\circ$ & -0.079 & $\pm 0.041$ & $\pm 0.013$ & -0.060 & $\pm 0.040$ & $\pm 0.015$\\ \hline 
62.5$^\circ$ &  0.023 & $\pm 0.045$ & $\pm 0.015$ &  0.000 & $\pm 0.044$ & $\pm 0.018$\\ \hline 
67.5$^\circ$ & -0.048 & $\pm 0.049$ & $\pm 0.016$ & -0.059 & $\pm 0.048$ & $\pm 0.022$\\ \hline 
72.5$^\circ$ &  0.037 & $\pm 0.052$ & $\pm 0.016$ & -0.081 & $\pm 0.050$ & $\pm 0.027$\\ \hline 
77.5$^\circ$ &  0.012 & $\pm 0.055$ & $\pm 0.016$ &  0.041 & $\pm 0.055$ & $\pm 0.031$\\ \hline 
82.5$^\circ$ & -0.038 & $\pm 0.059$ & $\pm 0.015$ &  0.014 & $\pm 0.057$ & $\pm 0.037$\\ \hline 
87.5$^\circ$ &  0.006 & $\pm 0.058$ & $\pm 0.014$ & -0.045 & $\pm 0.060$ & $\pm 0.042$\\ \hline 
\hline
\end{longtable}

\end{document}